\documentclass{article}
\usepackage{float}
\usepackage{graphicx}
\usepackage{amsmath,amssymb,amsthm}
\usepackage[colorlinks=true,linkcolor=blue,citecolor=blue,urlcolor=blue]{hyperref}

\title{Stable soap bubble clusters with multiple torus bubbles: getting a bit more exotic}
\author{Fabrice Delbary\\\href{mailto:fabrice.delbary@cnrs.fr}{fabrice.delbary@cnrs.fr}}
\date{}

\begin{document}
\maketitle
\section{Abstract}
Recently, numerical examples of stable soap bubble clusters with multiple torus bubbles have been presented \cite{delbary}. The geometry of these clusters is based on the Platonic solids whose vertices have valence $3$ (in order to fulfill Plateau's laws \cite{almgren-taylor,plateau}): the tetrahedron, the cube, the dodecahedron. The clusters respectively contain a bubble of genus $3, 5, 11$. The construction is quite generic and can be used with any convex polyhedron. If stable, the cluster obtained using a polyhedron with $n$ faces has $3n+2$ bubbles and one of these bubbles has genus $n-1$. We propose here to show that is it possible to get stable soap bubble clusters with multiple torus bubbles using a geometry based on prisms and Archimedean solids \cite{coxeter_73,coxeter_99,cromwell} as well.
\section{Introduction}\label{intro}
In \cite{delbary}, examples of stable soap bubble clusters with multiple torus bubbles are obtained by numerical simulations using Surface Evolver \cite{brakke}. Each starting geometry is defined by a Platonic solid $P$ and four parameters $0<s_{\mathrm{i}}<s_{\mathrm{m}}<s_{\mathrm{e}}<1$ and $0<s_{\mathrm{f}}<1$. The faces of $P$ define the outer interfaces of the outer bubbles. Three polyhedra $P_{\mathrm{i}},P_{\mathrm{m}},P_{\mathrm{e}}$ are defined by homothety of $P$ with respective ratios $s_{\mathrm{i}}<s_{\mathrm{m}}<s_{\mathrm{e}}$. The polyhedra $P_{\mathrm{e}},P$ define the outer bubbles. The polyhedron $P_{\mathrm{i}}$ defines the center bubble. The polyhedra $P_{\mathrm{i}},P_{\mathrm{e}}$ define the multiple torus bubble. The inner double bubbles are defined by right prisms whose bases are the faces of $P_{\mathrm{i}}$ scaled with a factor $s_{\mathrm{f}}$. The polyhedron $P_{\mathrm{m}}$ is used to define the interfaces shared by the inner double bubbles as the intersection with the right prisms.
\begin{figure}[H]
\centering
\includegraphics[scale=0.35]{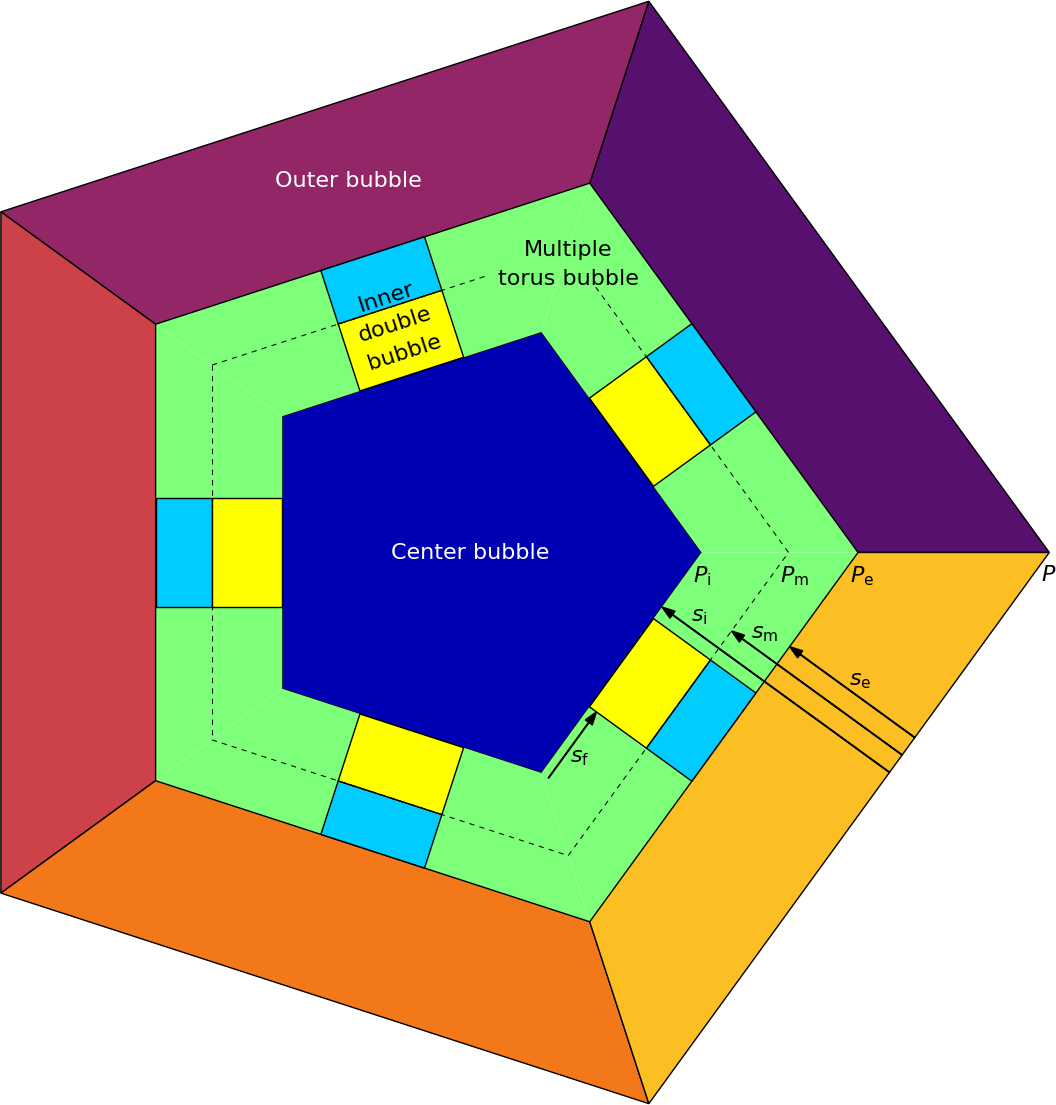}
\caption{$2$D scheme of the starting configuration.}
\label{init_conf}
\end{figure}
If the parameters are well-chosen, after area minimization under volume contraints (using Surface Evolver in our case), one gets a cluster which can be schematically represented in $2$D by the following figure.
\begin{figure}[H]
\centering
\includegraphics[scale=0.4]{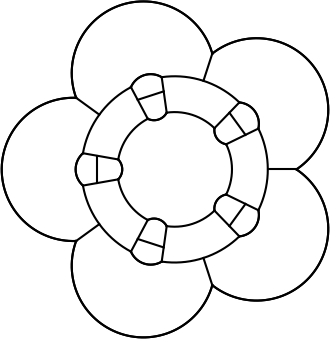}
\caption{$2$D schematic view of the stable cluster.}
\end{figure}
In \cite{delbary}, the parameters are chosen by trial and error: on the one hand, some parameters do not lead to a stable configuration with the same geometry, on the other hand, one has to choose parameters such that the interfaces dimensions of the stable cluster are large enough compared to the discretization. In this document, we keep the same construction, the same set of parameters $s_{\mathrm{i}},s_{\mathrm{m}},s_{\mathrm{e}},s_{\mathrm{f}}$ and proceed again by trial and error. Instead of Platonic solids, we define the starting geometries using prisms and Archimedean solids \cite{coxeter_73,coxeter_99,cromwell}. We work with polyhedra of unit edge length. We consider only the triangular, pentagonal, hexagonal prisms. Regarding the Archimedean solids, those whose vertices have valence $3$ are the truncated tetrahedron, the truncated cube, the truncated octahedron, the great rhombicuboctahedron (truncated cuboctahedron), the truncated dodecahedron, the truncated icosahedron, the great rhombicosidodecahedron (truncated icosidodecahedron). A set of coordinates for these polyhedra can be easily found by Wythoff construction \cite{coxeter_99,shephard}. Several discretizations, in terms of highest triangle edge length, are considered and once the stable configuration has been reached, the Hessian matrix lowest and highest eigenvalues are computed (hence its condition number too). For each configuration, a figure will show the whole cluster, the cluster without the outer bubbles, the center bubble with the inner double bubbles and the multiple torus bubble alone.
\section{Semiregular prisms}
For $n\geq3$, the semiregular $n$-prism has two regular $n$-gonal faces and $n$ square faces, $3n$ edges and $2n$ vertices. Vertices coordinates for the semiregular $n$-prism of unit edge length can be chosen as
\begin{equation}
\frac{1}{2\sin\frac{\pi}{n}}\begin{pmatrix}\cos\frac{2k\pi}{n}\\\sin\frac{2k\pi}{n}\\\pm\sin\frac{\pi}{n}\end{pmatrix}\quad,\quad k=0,\ldots,n-1.
\end{equation}
The bubble cluster constructed using an $n$-prism has $3n+8$ bubbles and the multiple torus bubble has genus $n+1$.
\subsection{Triangular prism}
The cluster has $17$ bubbles and the multiple torus bubble has genus $4$.
\begin{figure}[H]
\centering
\includegraphics[scale=0.17]{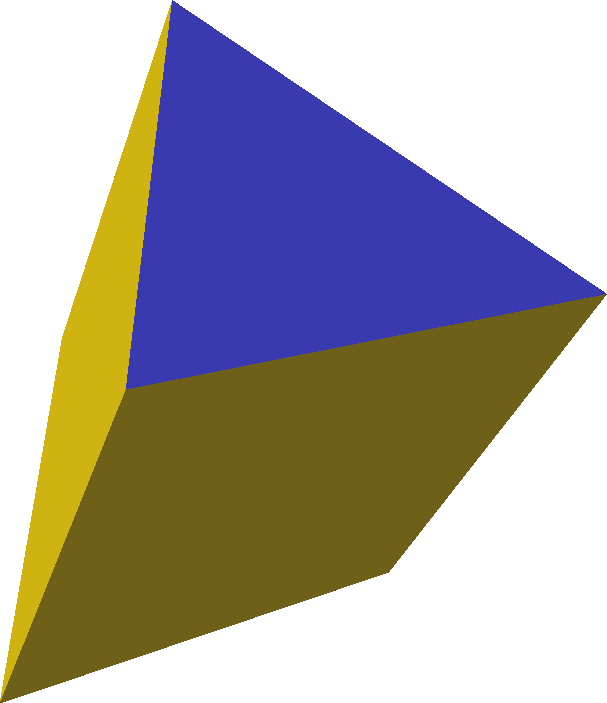}\includegraphics[scale=0.17]{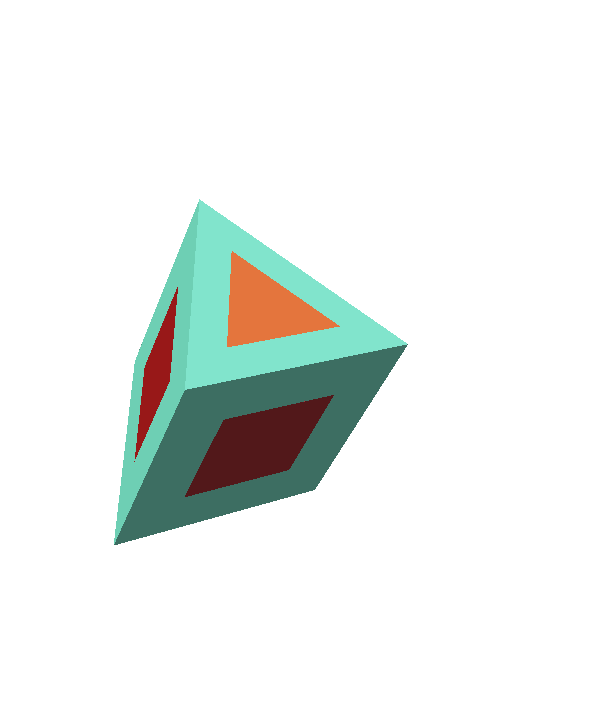}\hspace{-1.4cm}\includegraphics[scale=0.17]{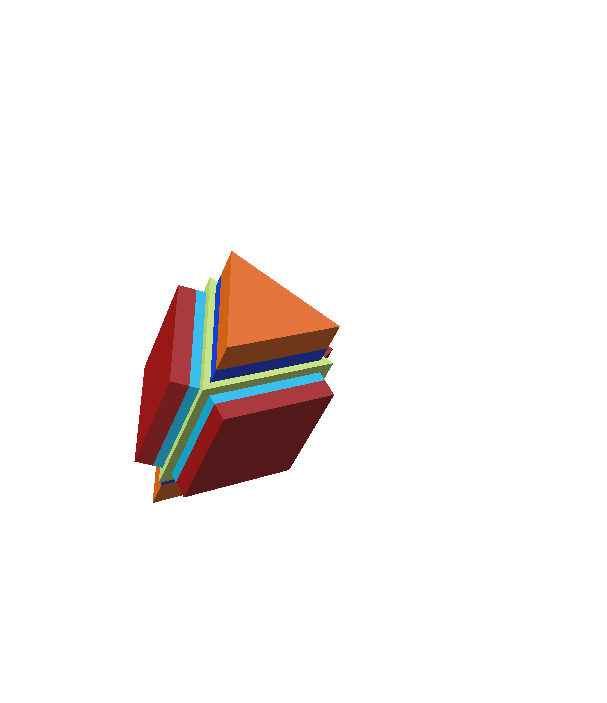}\hspace{-1.5cm}\includegraphics[scale=0.17]{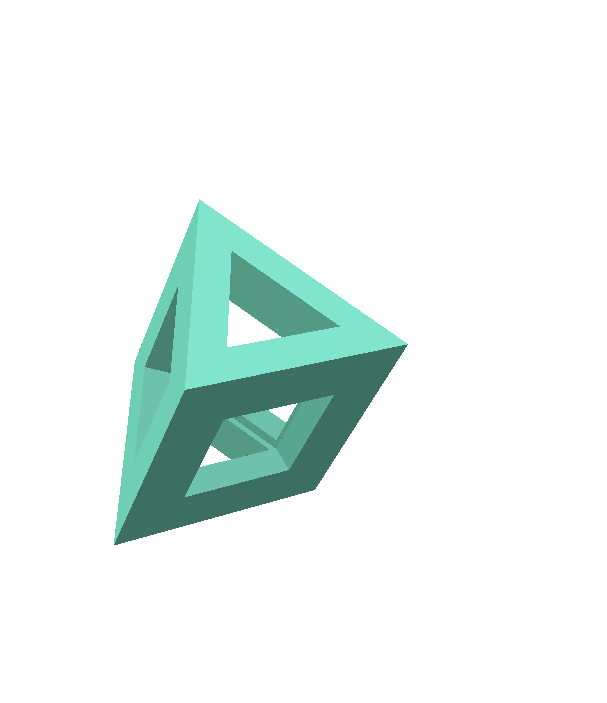}
\caption{Starting geometry. $s_{\mathrm{i}}=0.3\;,\;s_{\mathrm{m}}=0.37\;,\;s_{\mathrm{e}}=0.5\;,\;s_{\mathrm{f}}=0.85$.}
\end{figure}
\begin{figure}[H]
\centering
\includegraphics[scale=0.18]{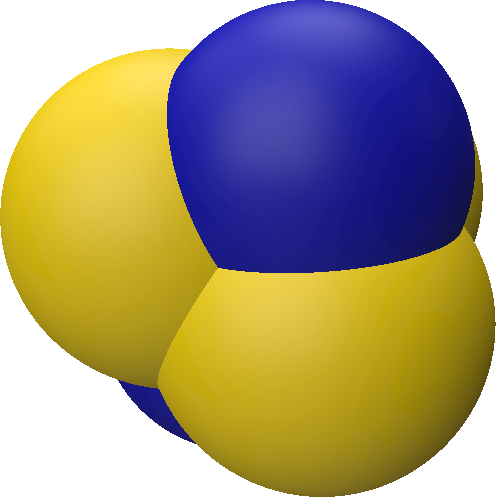}\includegraphics[scale=0.18]{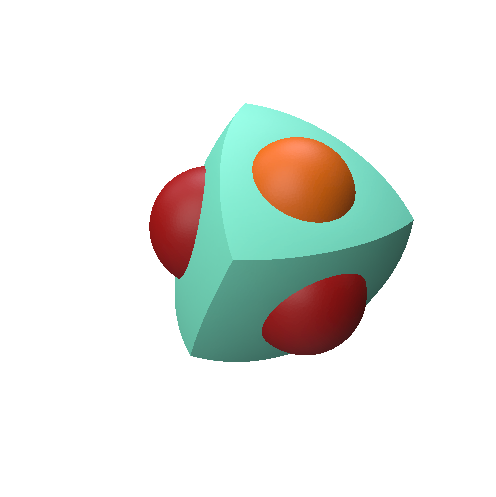}\hspace{-0.5cm}\includegraphics[scale=0.18]{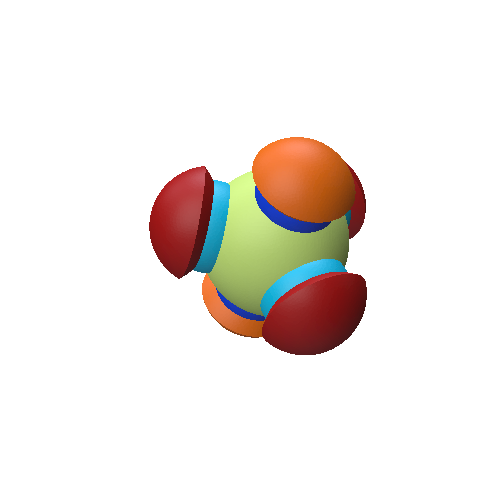}\hspace{-0.7cm}\includegraphics[scale=0.18]{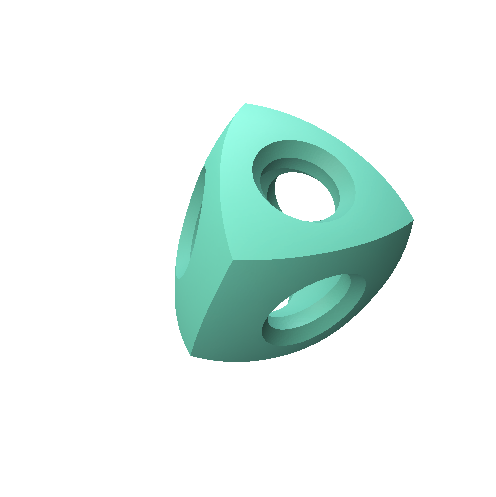}
\caption{Stable cluster. Total area: $4.96$.}
\end{figure}
\begin{table}[H]
\centering
\begin{tabular}{|c|c|c|}
\hline
Discretization & Lowest eigenvalue & Highest eigenvalue\\
\hline
$10^{-2}$ & $6.4\times10^{-6}$ & $12$\\
\hline
$8\times10^{-3}$ & $4.1\times10^{-6}$ & $11$\\
\hline
$4\times10^{-3}$ & $4.5\times10^{-7}$ & $14$\\
\hline
$2\times10^{-3}$ & $5.6\times10^{-8}$ & $13$\\
\hline
\end{tabular}
\caption{Hessian matrix eigenvalues for the different discretizations.}
\end{table}
\subsection{Pentagonal prism}
The cluster has $23$ bubbles and the multiple torus bubble has genus $6$.
\begin{figure}[H]
\centering
\includegraphics[scale=0.14]{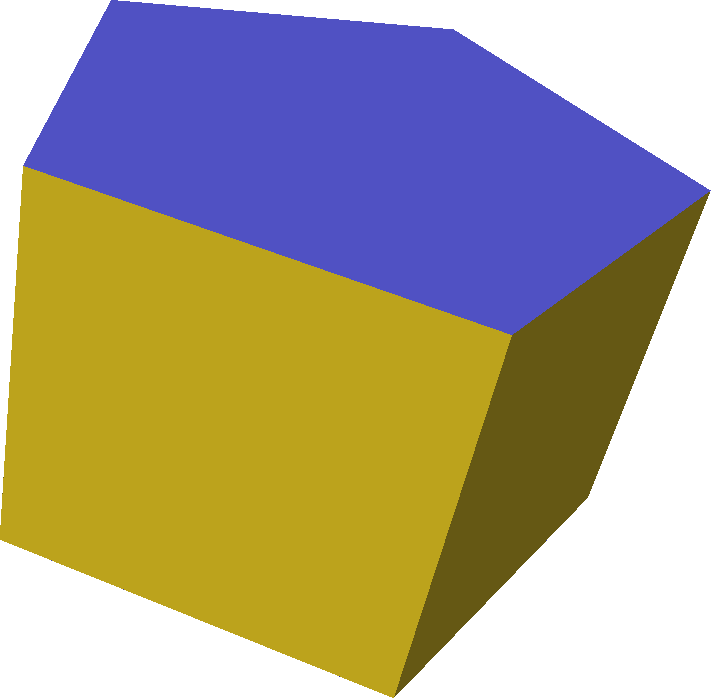}\includegraphics[scale=0.14]{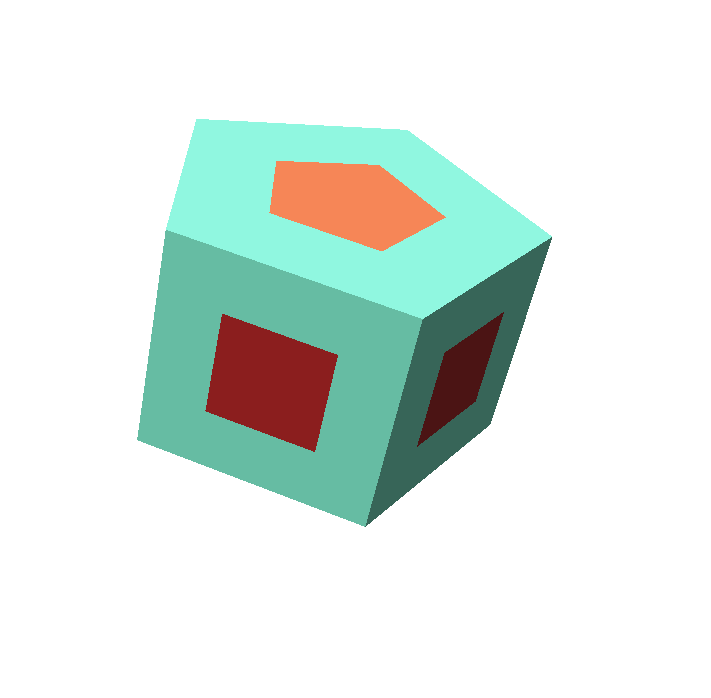}\hspace{-1cm}\includegraphics[scale=0.14]{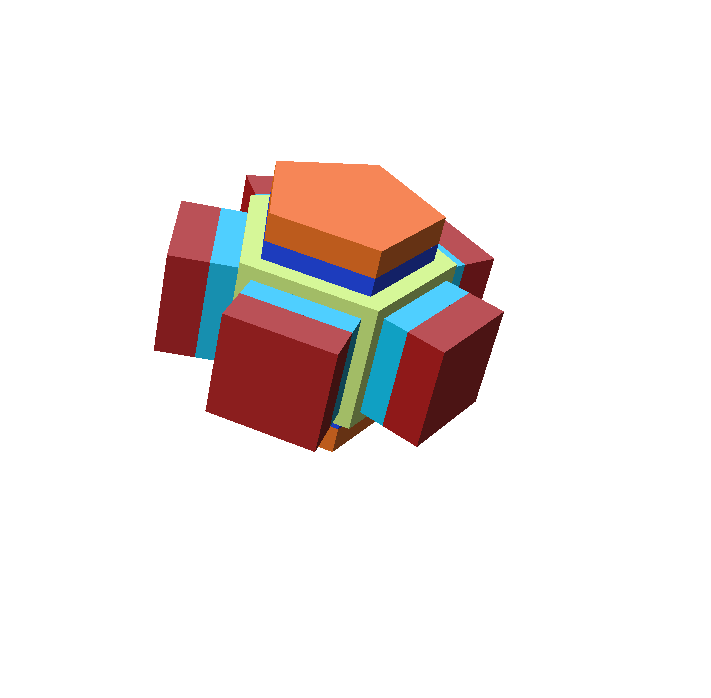}\hspace{-1cm}\includegraphics[scale=0.14]{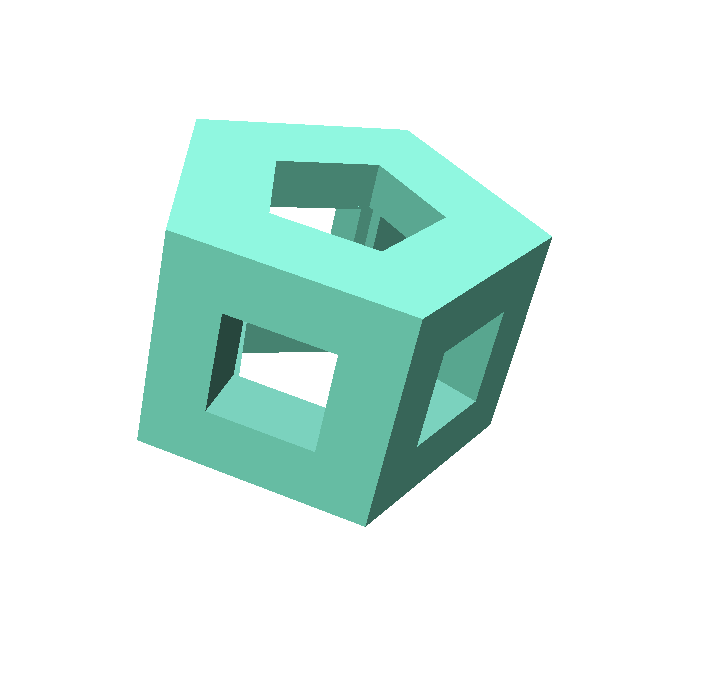}
\caption{Starting geometry. $s_{\mathrm{i}}=0.35\;,\;s_{\mathrm{m}}=0.45\;,\;s_{\mathrm{e}}=0.6\;,\;s_{\mathrm{f}}=0.8$.}
\end{figure}
\begin{figure}[H]
\centering
\includegraphics[scale=0.14]{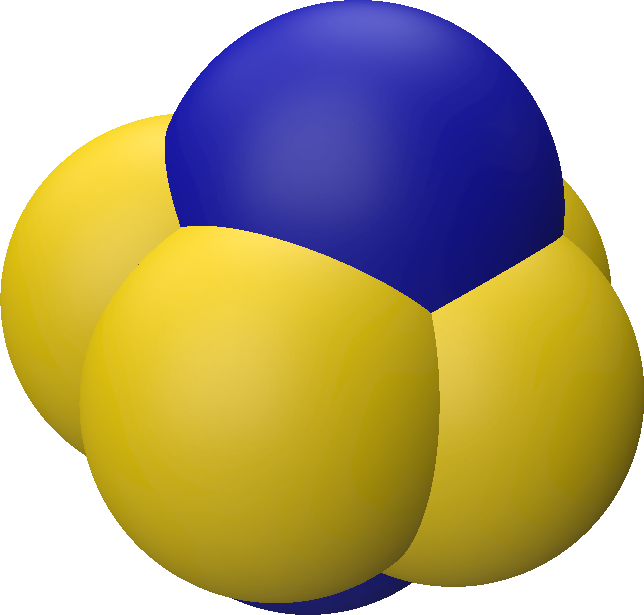}\includegraphics[scale=0.14]{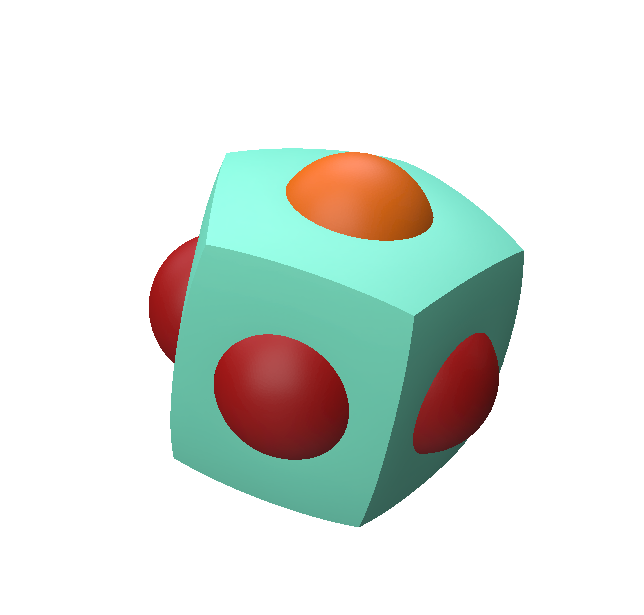}\hspace{-0.5cm}\includegraphics[scale=0.14]{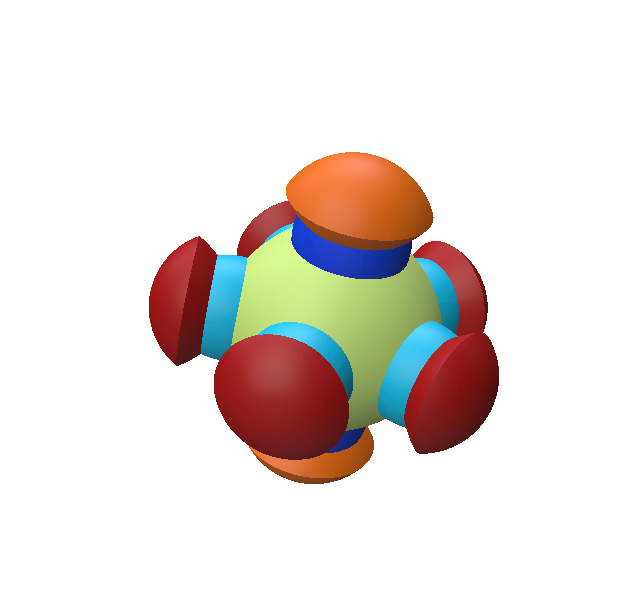}\hspace{-0.7cm}\includegraphics[scale=0.14]{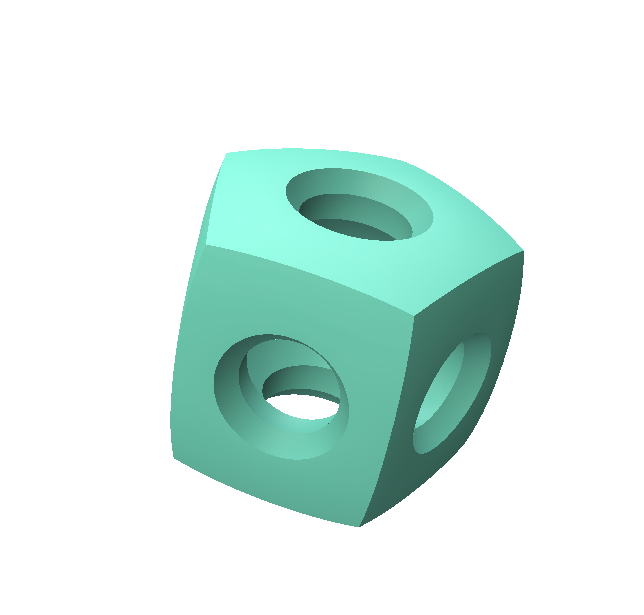}
\caption{Stable cluster. Total area: $14.05$.}
\end{figure}
\begin{table}[H]
\centering
\begin{tabular}{|c|c|c|}
\hline
Discretization & Lowest eigenvalue & Highest eigenvalue\\
\hline
$2\times10^{-2}$ & $2.0\times10^{-5}$ & $10$\\
\hline
$8\times10^{-3}$ & $1.5\times10^{-6}$ & $11$\\
\hline
$4\times10^{-3}$ & $1.7\times10^{-7}$ & $13$\\
\hline
$3\times10^{-3}$ & $9.2\times10^{-8}$ & $11$\\
\hline
\end{tabular}
\caption{Hessian matrix eigenvalues for the different discretizations.}
\end{table}
\subsection{Hexagonal prism}
The cluster has $26$ bubbles and the multiple torus bubble has genus $7$.
\begin{figure}[H]
\centering
\includegraphics[scale=0.12]{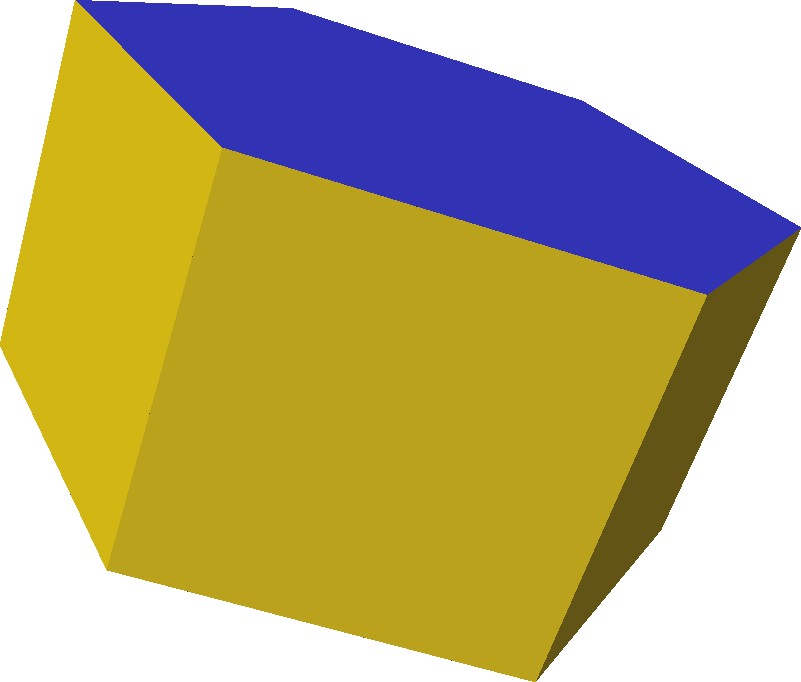}\includegraphics[scale=0.12]{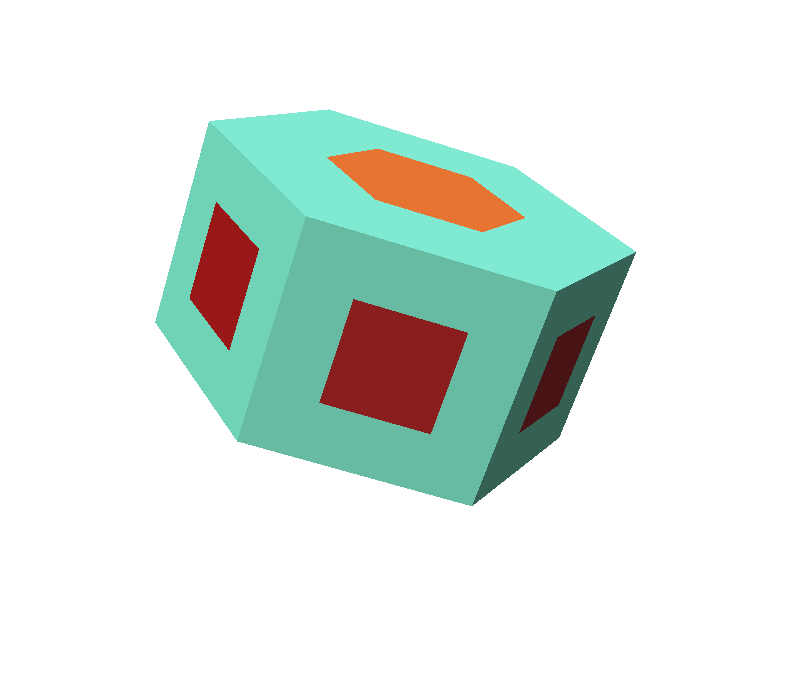}\hspace{-1cm}\includegraphics[scale=0.12]{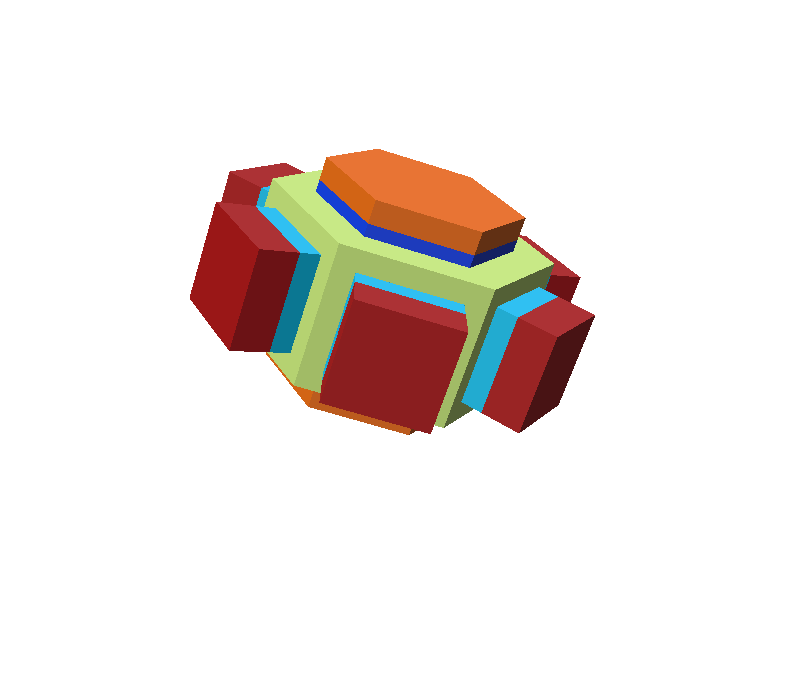}\hspace{-1cm}\includegraphics[scale=0.12]{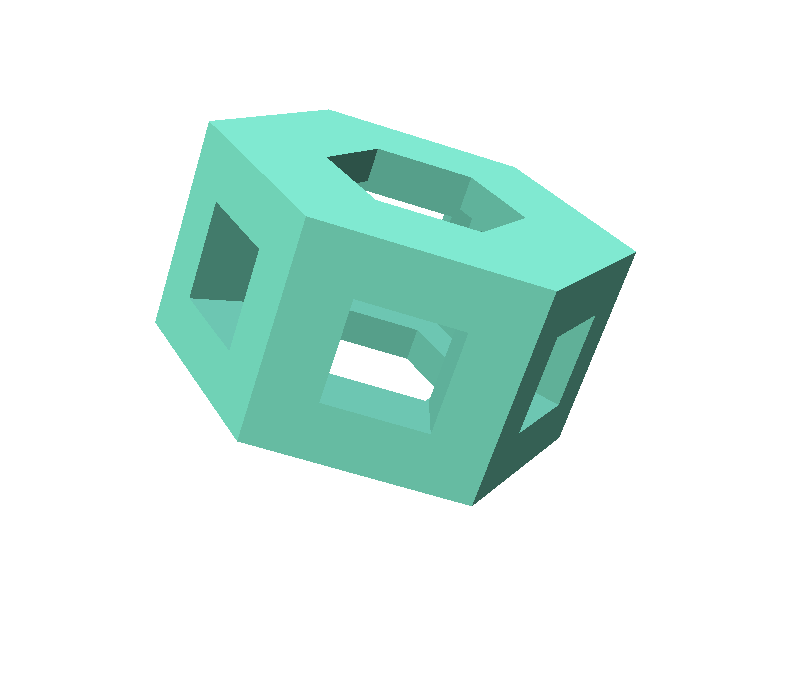}
\caption{Starting geometry. $s_{\mathrm{i}}=0.4\;,\;s_{\mathrm{m}}=0.47\;,\;s_{\mathrm{e}}=0.6\;,\;s_{\mathrm{f}}=0.7$.}
\end{figure}
\begin{figure}[H]
\centering
\includegraphics[scale=0.13]{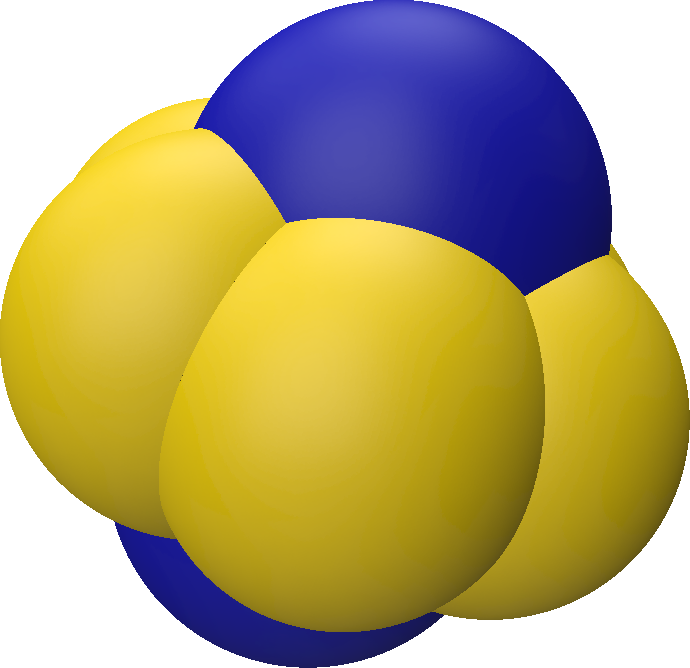}\includegraphics[scale=0.13]{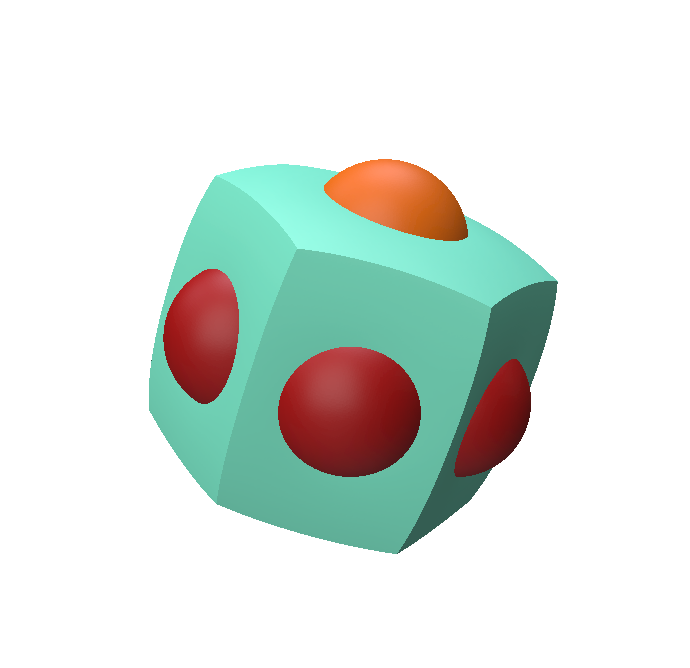}\hspace{-0.5cm}\includegraphics[scale=0.13]{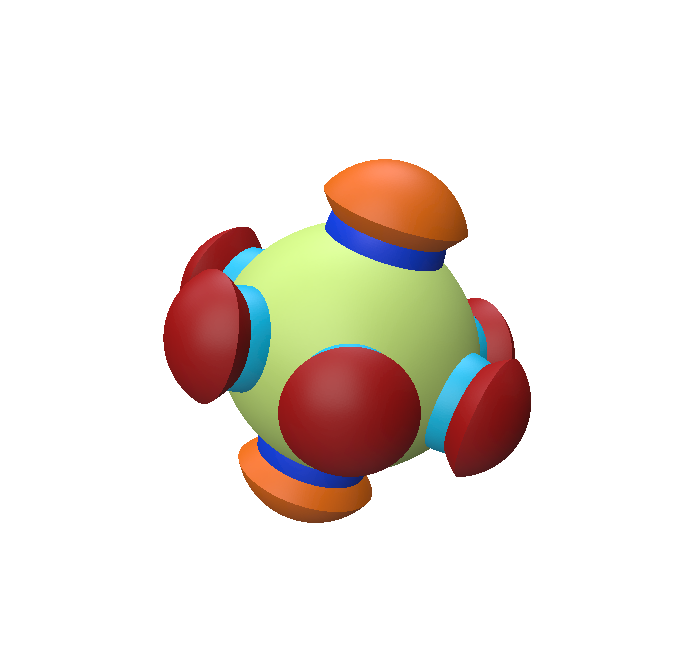}\hspace{-0.7cm}\includegraphics[scale=0.13]{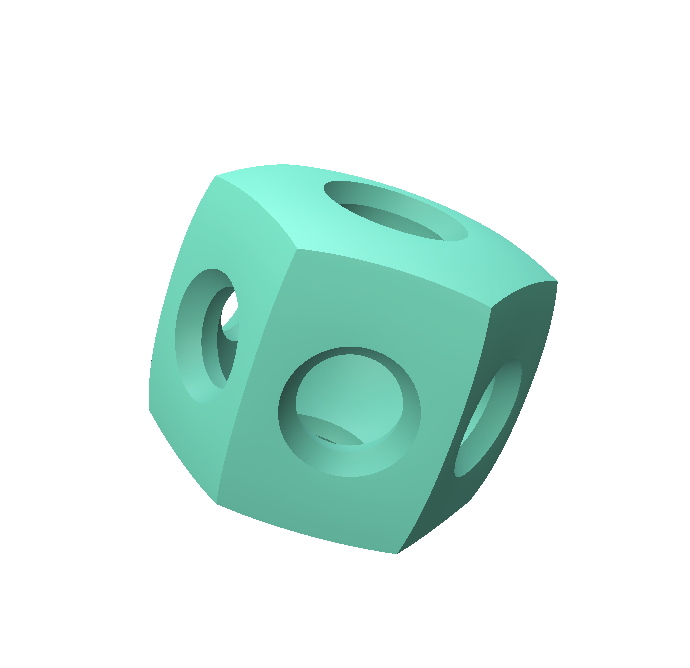}
\caption{Stable cluster. Total area: $18.90$.}
\end{figure}
\begin{table}[H]
\centering
\begin{tabular}{|c|c|c|}
\hline
Discretization & Lowest eigenvalue & Highest eigenvalue\\
\hline
$2\times10^{-2}$ & $1.5\times10^{-5}$ & $11$\\
\hline
$10^{-2}$ & $1.8\times10^{-6}$ & $11$\\
\hline
$8\times10^{-3}$ & $1.1\times10^{-6}$ & $11$\\
\hline
$4\times10^{-3}$ & $1.2\times10^{-7}$ & $13$\\
\hline
\end{tabular}
\caption{Hessian matrix eigenvalues for the different discretizations.}
\end{table}
\subsection{Prisms with more sides}
No test has been performed for prisms with more sides. If one keeps the geometry of a semiregular prism, as the number of sides increases, the bubbles associated with the two $n$-gons become much larger than the side ones. For this reason, it might be wise to consider non semiregular prisms and increase their height in order to have larger side bubbles. At the same time, it could be worthwhile to consider scaling factors $s_{\mathrm{f}}$ depending on the kind of face. The number $s_{\mathrm{m}}$ could also be defined per face. All these considerations hold for the Archimedean solids treated in the following as well. One can work with irregular convex polyhedra having the same combinatorial structure in order to change the size of faces (e.g. by changing the truncation), hence the volumes of the corresponding bubbles, and/or take a definition of the inner double bubbles depending on which kind of face they are linked to. In fact, in the case of more general convex polyhedra, one might need a fine tuning of all the bubble volumes. To this end, the homothety center can be placed anywhere inside the polyhedra and the inner double bubbles can be defined as frusta instead of prisms. Compared to figure \ref{init_conf}, one would get this kind of starting configuration.
\begin{figure}[H]
\centering
\includegraphics[scale=0.15]{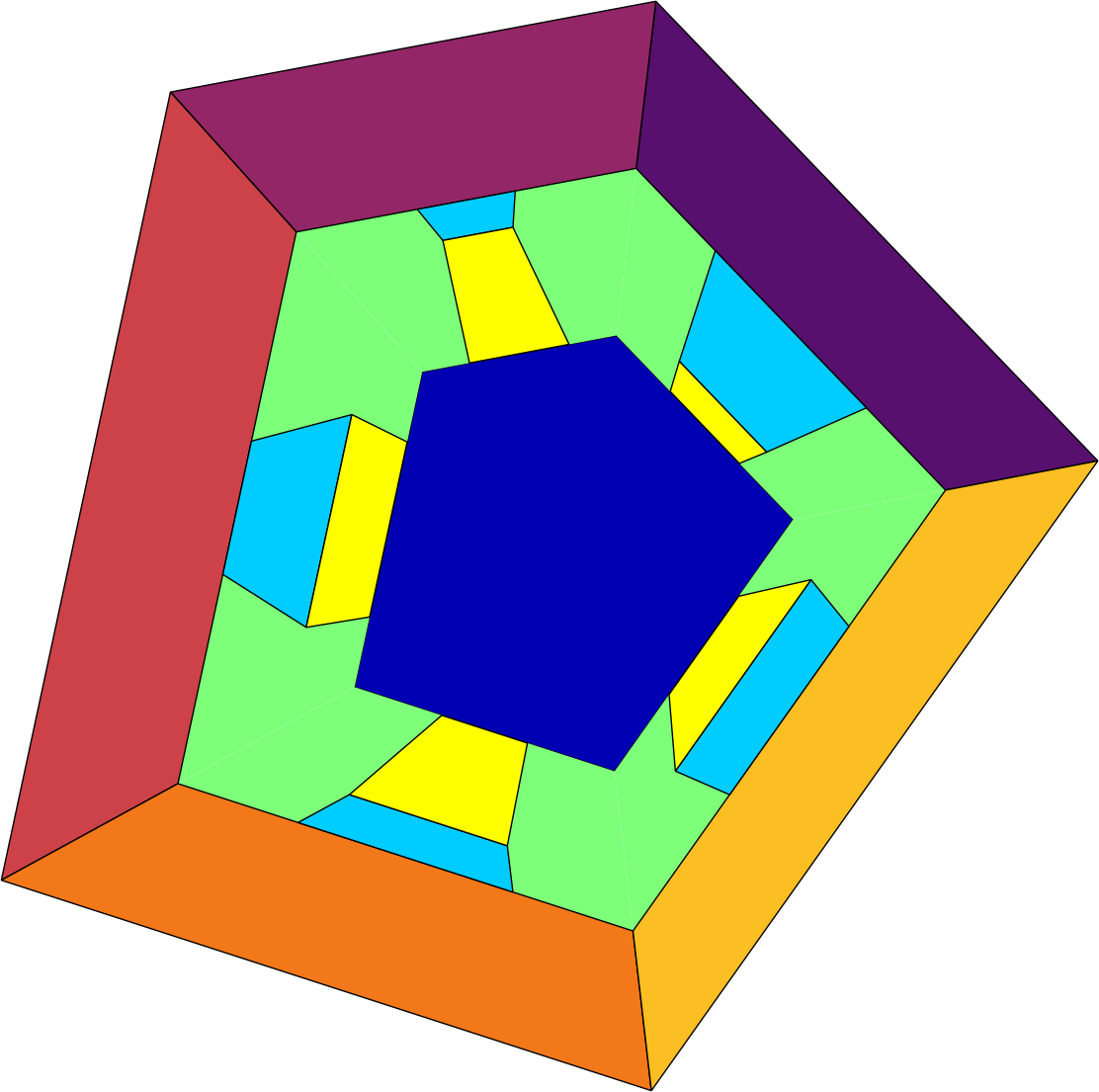}
\caption{$2$D scheme of a more general starting configuration.}
\label{modified_initial_configuration}
\end{figure}
\subsection{Truncated tetrahedron}
The truncated tetrahedron has $4$ triangular faces and $4$ hexagonal faces, $18$ edges and $12$ vertices. The cluster has $26$ bubbles and the multiple torus bubble has genus $7$. Vertices coordinates of the truncated tetrahedron of unit edge length can be chosen as the permutations of
\begin{equation}
\frac{1}{2\sqrt{2}}\begin{pmatrix}\pm3\\\pm1\\\pm1\end{pmatrix}.
\end{equation}
\begin{figure}[H]
\centering
\includegraphics[scale=0.13]{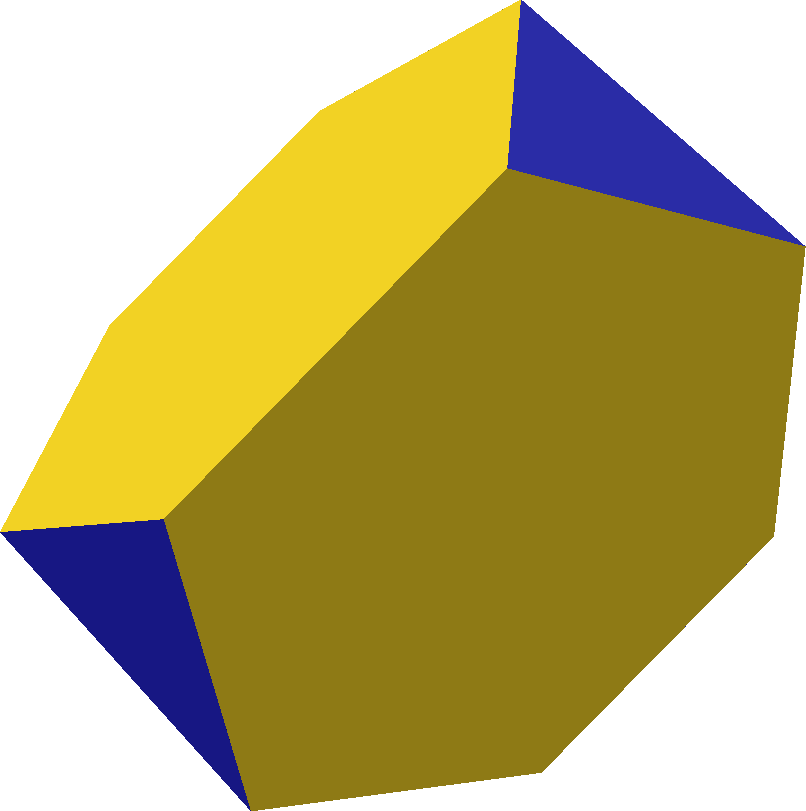}\includegraphics[scale=0.13]{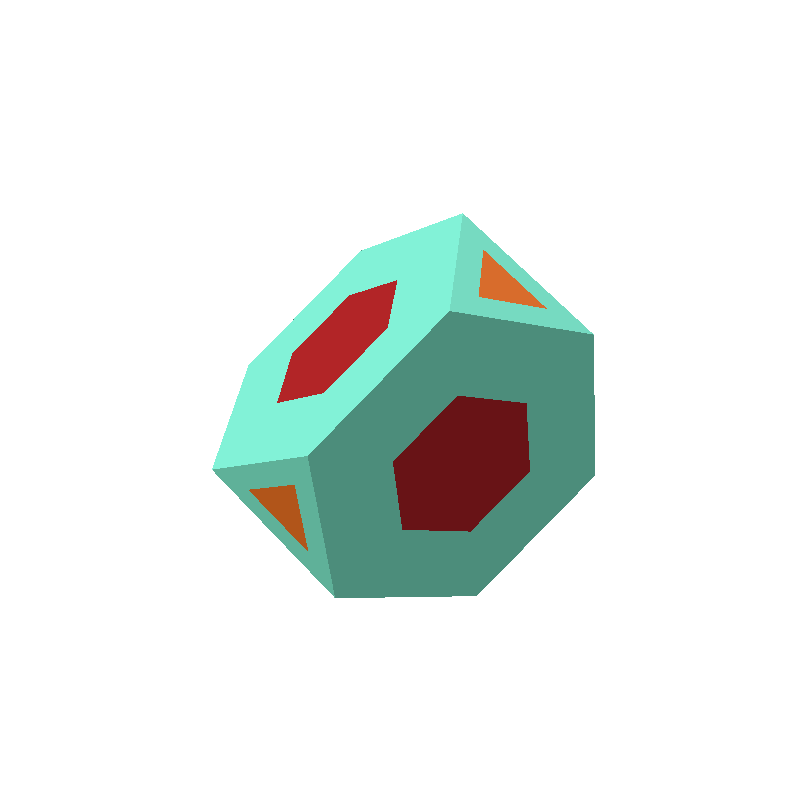}\hspace{-1.5cm}\includegraphics[scale=0.13]{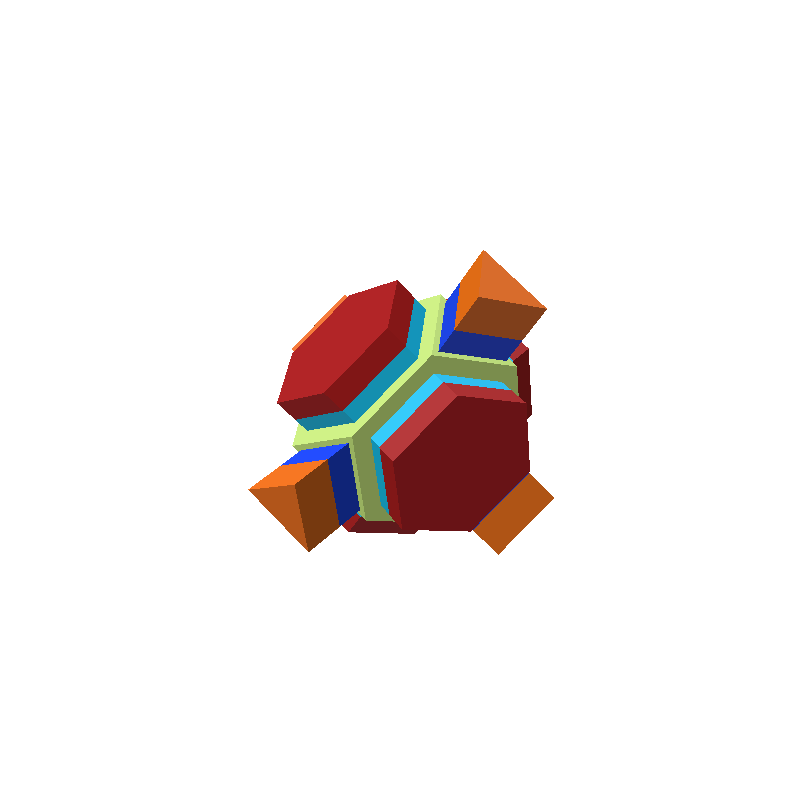}\hspace{-1.5cm}\includegraphics[scale=0.13]{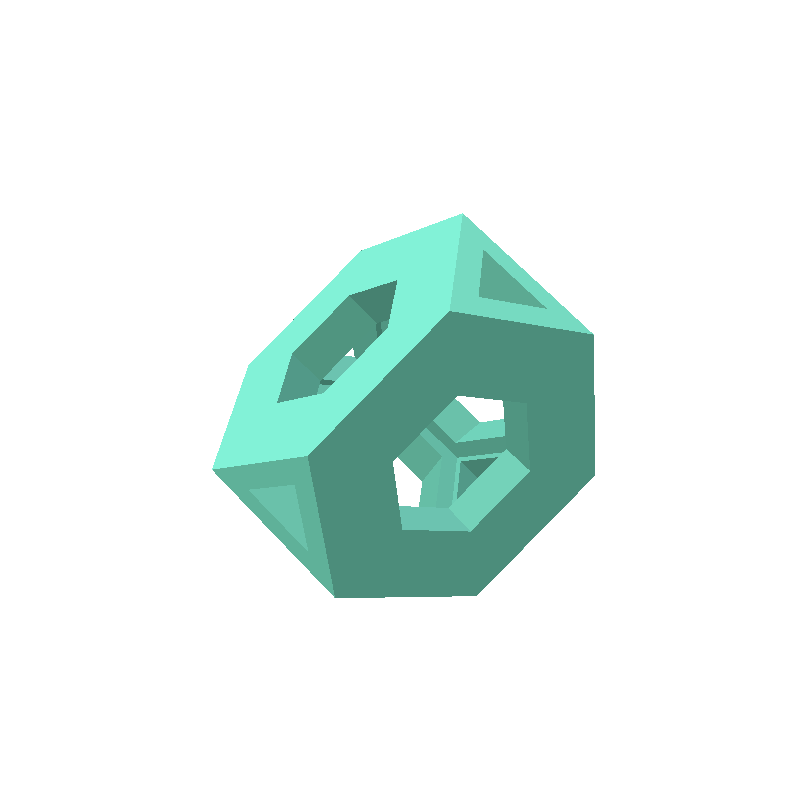}
\caption{Starting geometry. $s_{\mathrm{i}}=0.3\;,\;s_{\mathrm{m}}=0.38\;,\;s_{\mathrm{e}}=0.5\;,\;s_{\mathrm{f}}=0.8$.}
\end{figure}
\begin{figure}[H]
\centering
\includegraphics[scale=0.15]{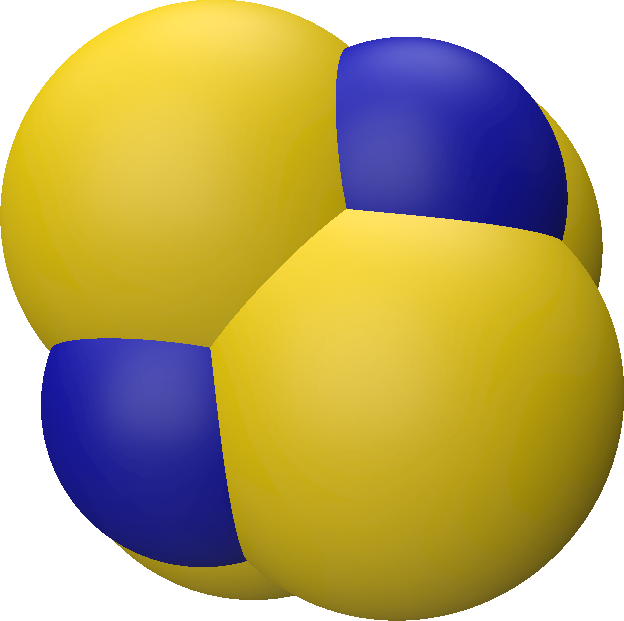}\includegraphics[scale=0.15]{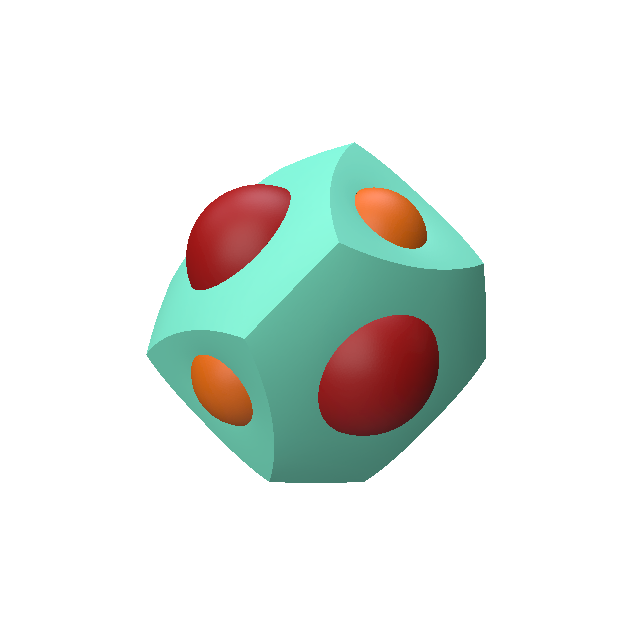}\hspace{-0.5cm}\includegraphics[scale=0.15]{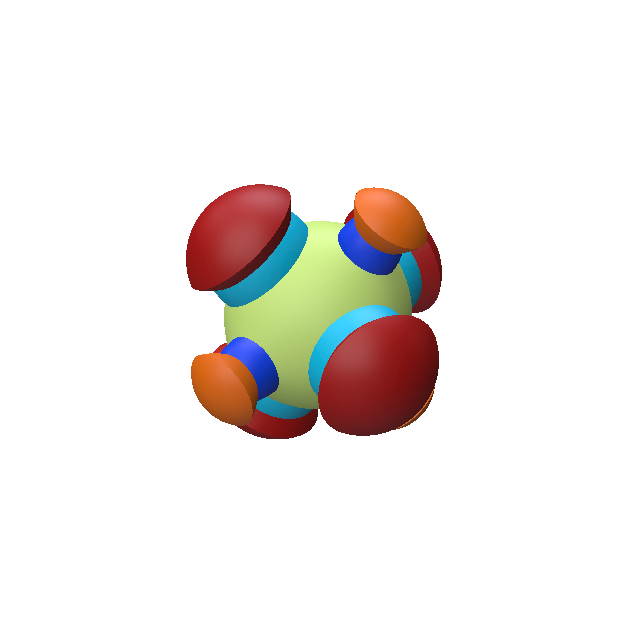}\hspace{-0.7cm}\includegraphics[scale=0.15]{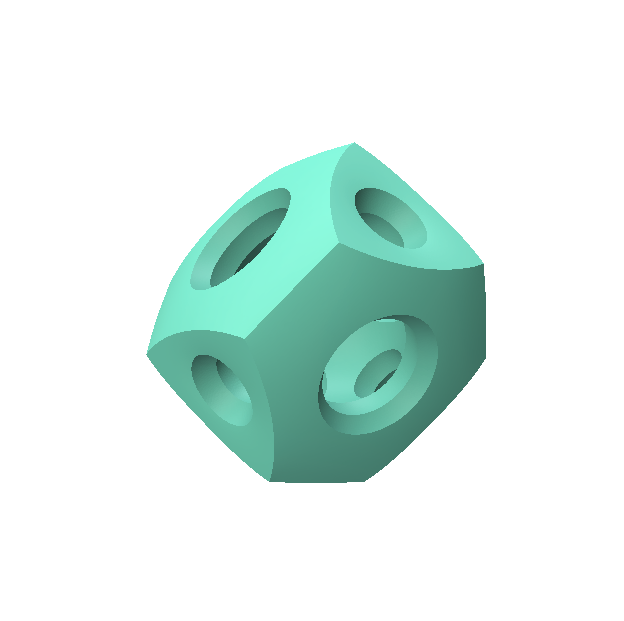}
\caption{Stable cluster. Total area: $17.90$.}
\end{figure}
\begin{table}[H]
\centering
\begin{tabular}{|c|c|c|}
\hline
Discretization & Lowest eigenvalue & Highest eigenvalue\\
\hline
$2\times10^{-2}$ & $1.6\times10^{-5}$ & $11$\\
\hline
$8\times10^{-3}$ & $8.7\times10^{-7}$ & $15$\\
\hline
$4\times10^{-3}$ & $1.1\times10^{-7}$ & $15$\\
\hline
\end{tabular}
\caption{Hessian matrix eigenvalues for the different discretizations.}
\end{table}
\subsection{Truncated cube}
The truncated cube has $8$ triangular faces and $6$ octagonal faces, $36$ edges and $24$ vertices. The cluster has $44$ bubbles and the multiple torus bubble has genus $13$. Vertices coordinates of the truncated cube of unit edge length can be chosen as the permutations of
\begin{equation}
\frac{1}{2}\begin{pmatrix}\pm1\\\pm(1+\sqrt{2})\\\pm(1+\sqrt{2})\end{pmatrix}.
\end{equation}
\begin{figure}[H]
\centering
\includegraphics[scale=0.12]{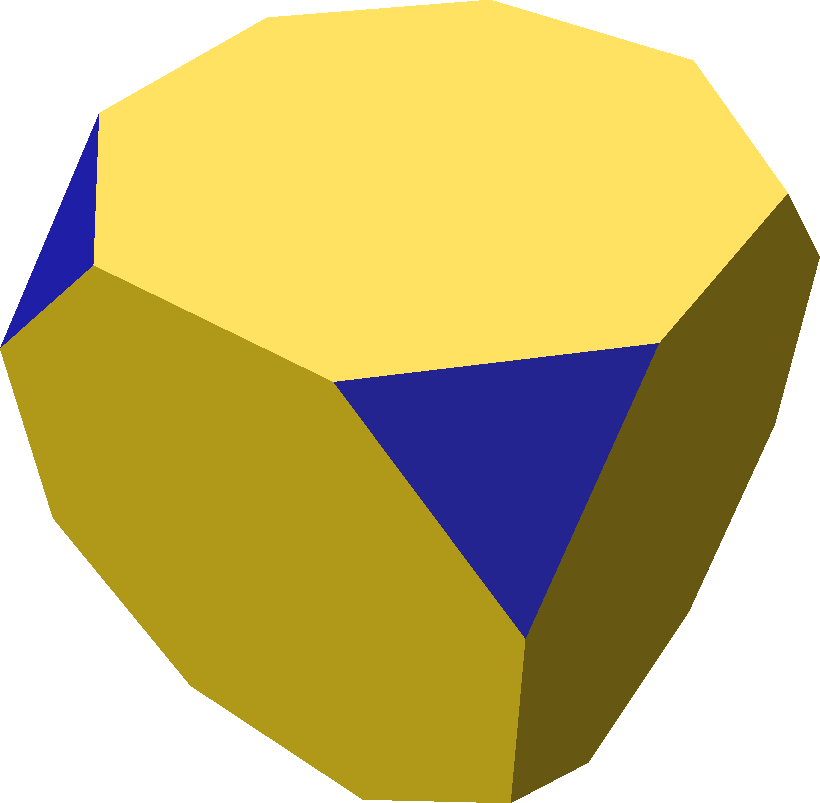}\hspace{-0.4cm}\includegraphics[scale=0.12]{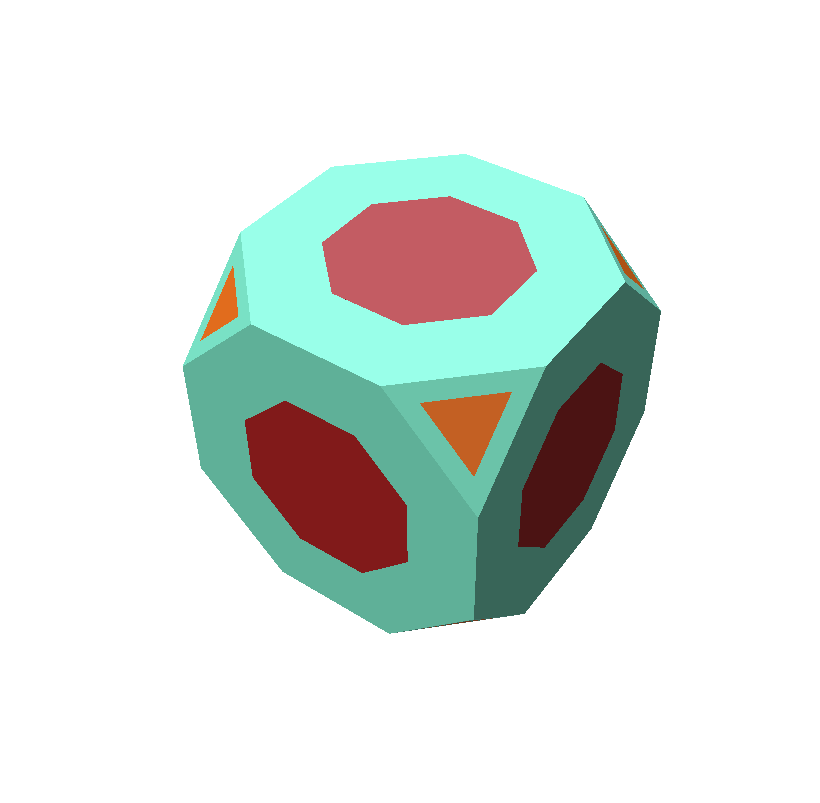}\hspace{-1.2cm}\includegraphics[scale=0.12]{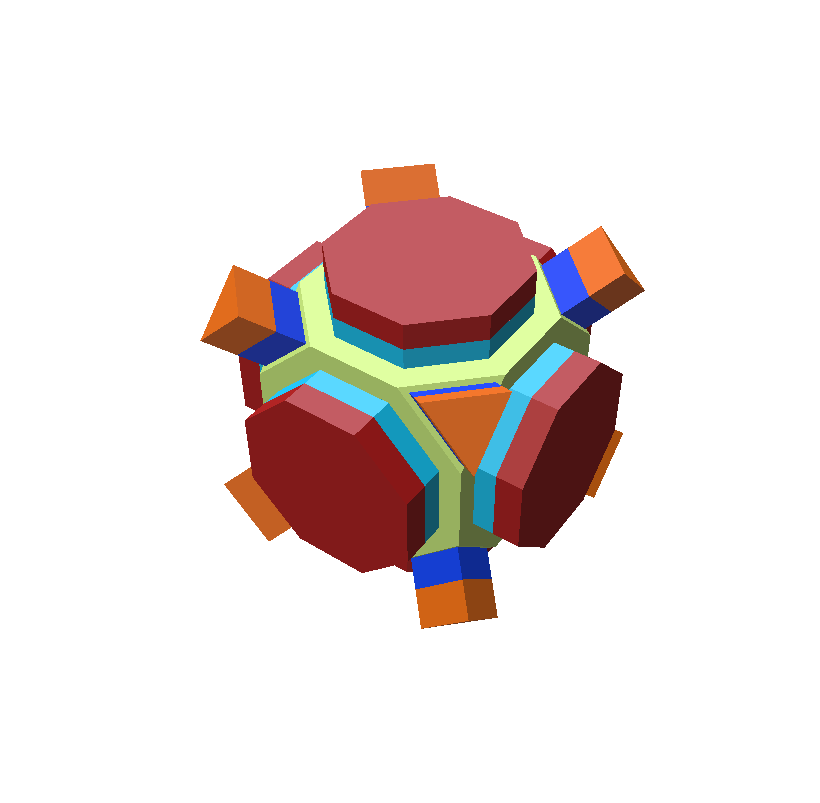}\hspace{-1.2cm}\includegraphics[scale=0.12]{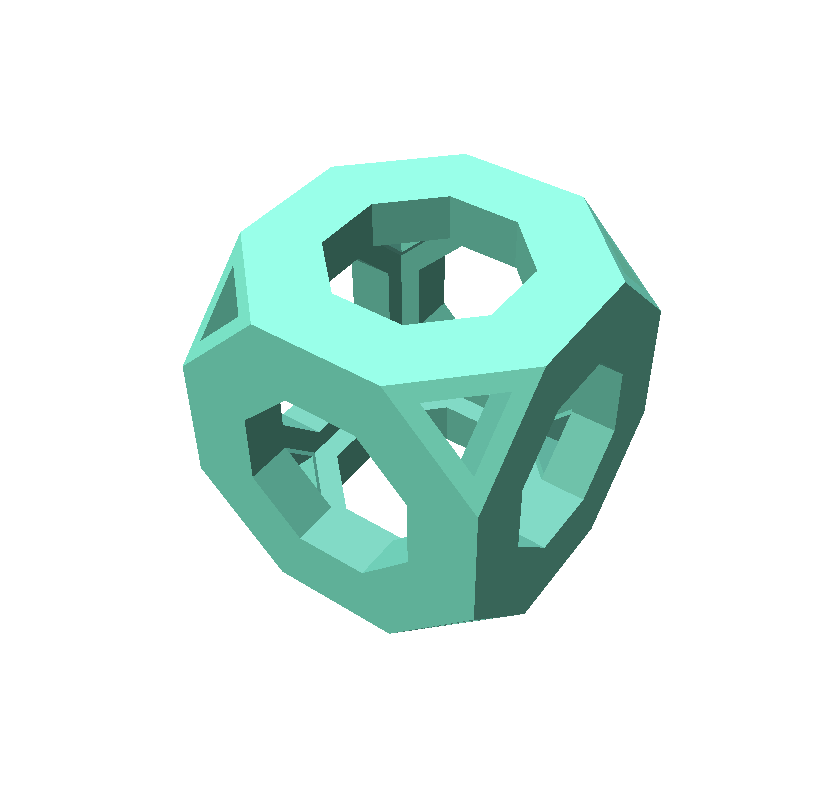}
\caption{Starting geometry. $s_{\mathrm{i}}=0.42\;,\;s_{\mathrm{m}}=0.5\;,\;s_{\mathrm{e}}=0.6\;,\;s_{\mathrm{f}}=0.8$.}
\end{figure}
\begin{figure}[H]
\centering
\includegraphics[scale=0.13]{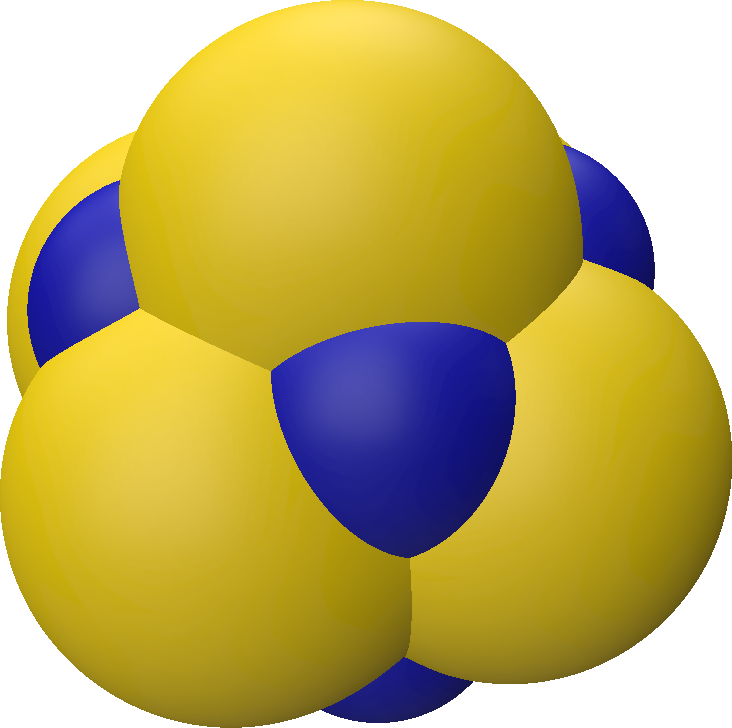}\includegraphics[scale=0.13]{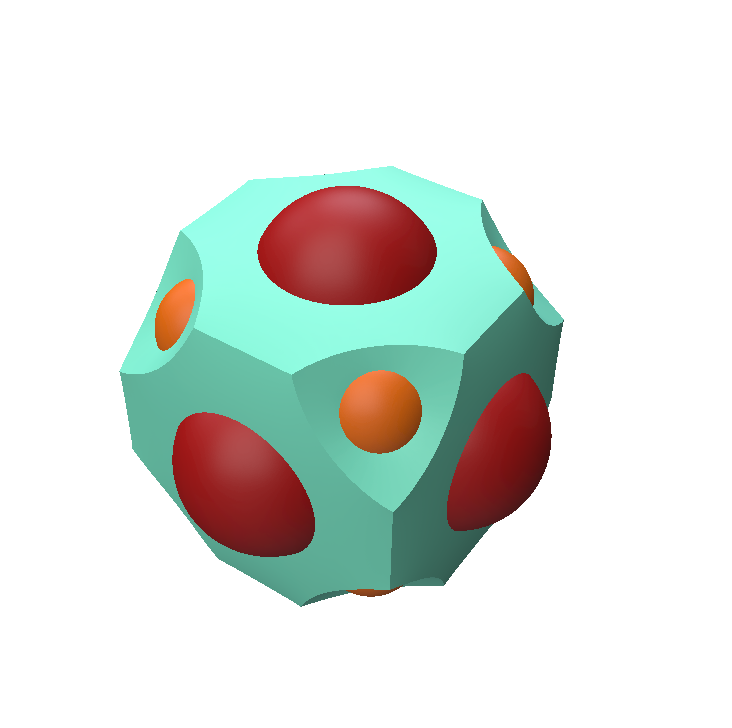}\hspace{-1cm}\includegraphics[scale=0.13]{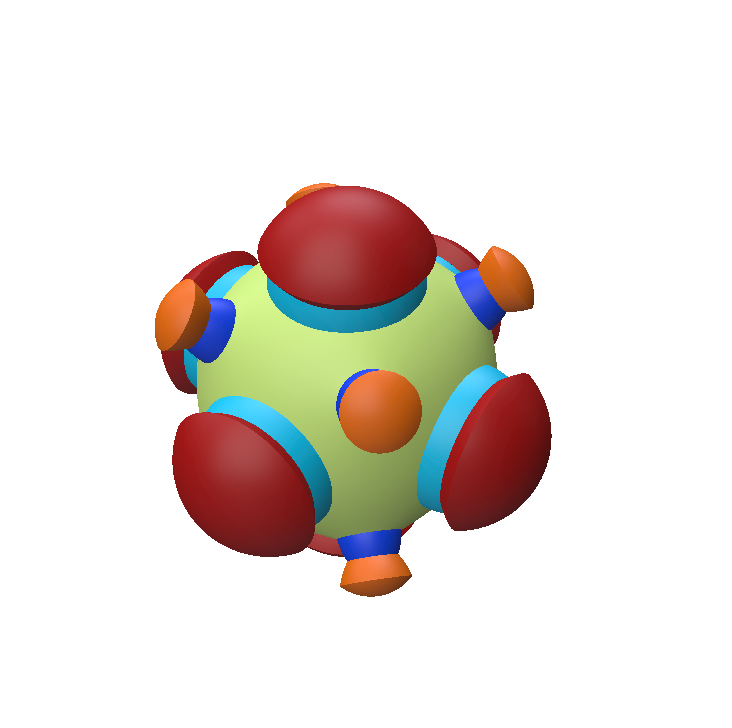}\hspace{-1cm}\includegraphics[scale=0.13]{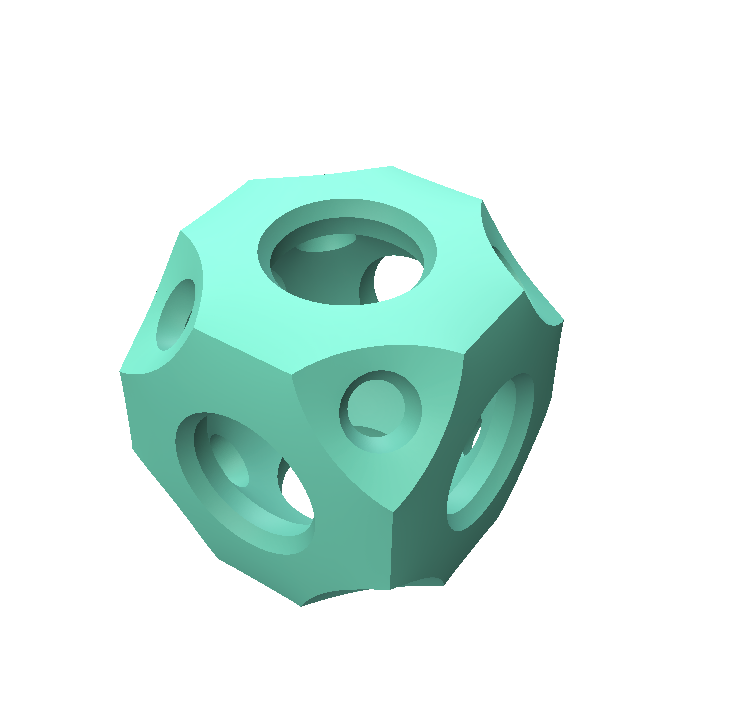}
\caption{Stable cluster. Total area: $61.46$.}
\end{figure}
\begin{table}[H]
\centering
\begin{tabular}{|c|c|c|}
\hline
Discretization & Lowest eigenvalue & Highest eigenvalue\\
\hline
$3\times10^{-2}$ & $2.0\times10^{-5}$ & $13$\\
\hline
$10^{-2}$ & $7.6\times10^{-7}$ & $15$\\
\hline
$8\times10^{-3}$ & $3.4\times10^{-7}$ & $17$\\
\hline
\end{tabular}
\caption{Hessian matrix eigenvalues for the different discretizations.}
\end{table}
\subsection{Truncated octahedron}
The truncated octahedron has $6$ square faces and $8$ hexagonal faces, $36$ edges and $24$ vertices. The cluster has $44$ bubbles and the multiple torus bubble has genus $13$. Vertices coordinates of the truncated octahedron of unit edge length can be chosen as the permutations of
\begin{equation}
\begin{pmatrix}\pm\sqrt{2}\\\pm\frac{1}{\sqrt{2}}\\0\end{pmatrix}.
\end{equation}
\begin{figure}[H]
\centering
\includegraphics[scale=0.12]{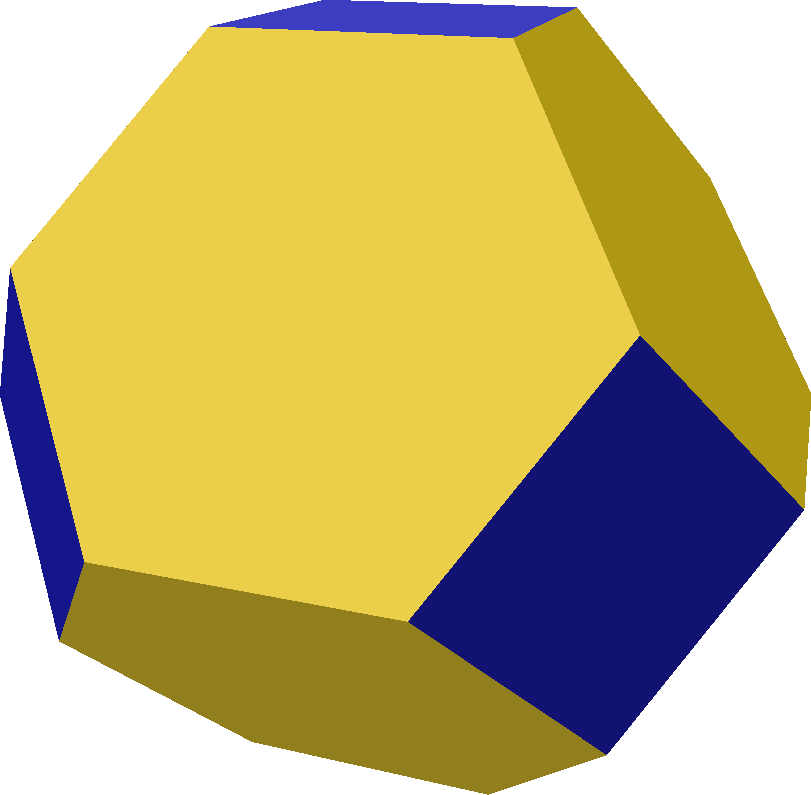}\hspace{-0.4cm}\includegraphics[scale=0.12]{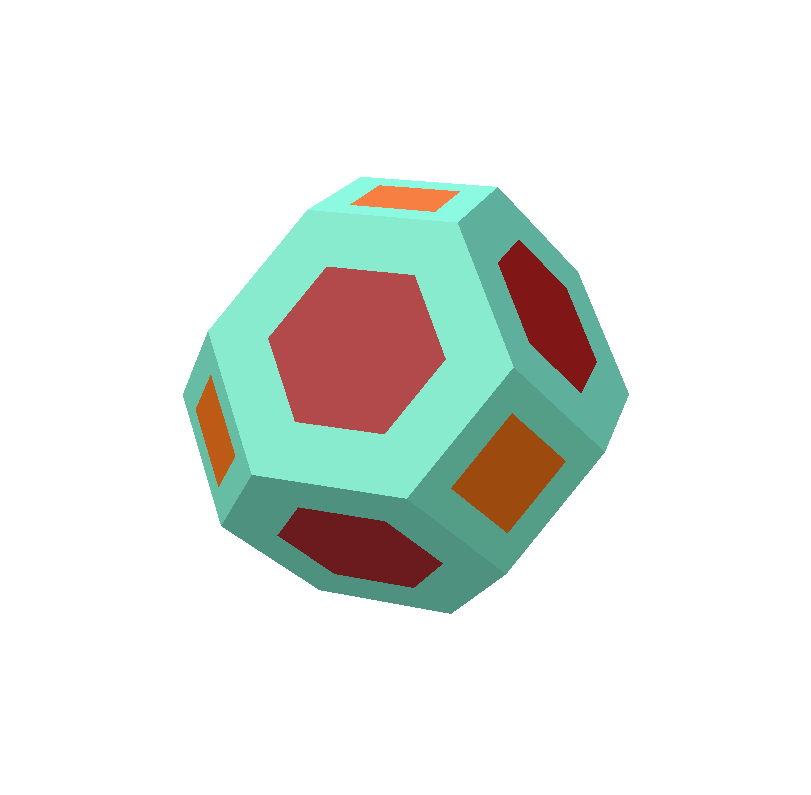}\hspace{-1.2cm}\includegraphics[scale=0.12]{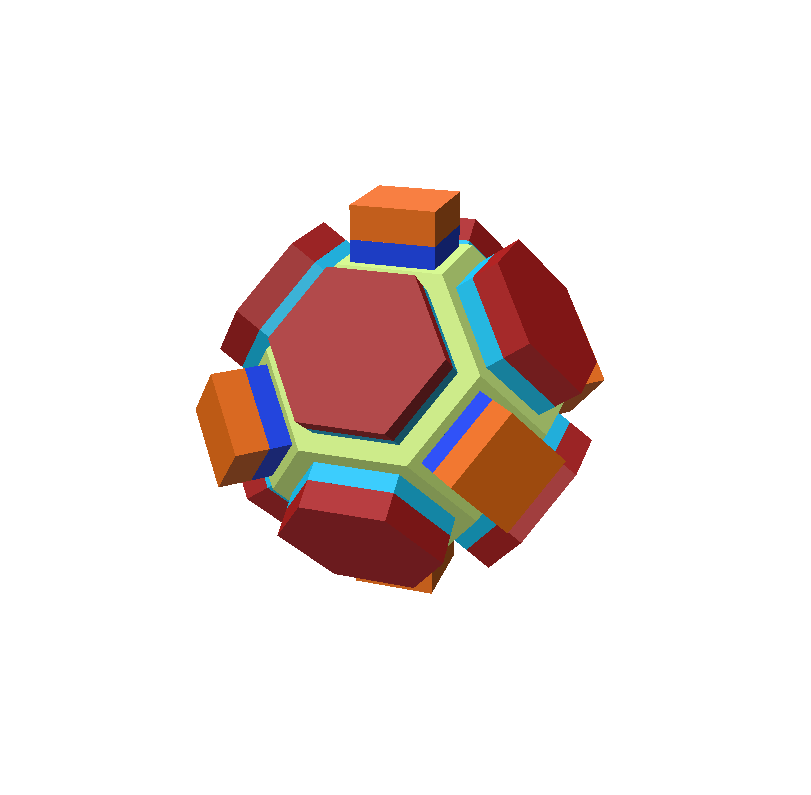}\hspace{-1.2cm}\includegraphics[scale=0.12]{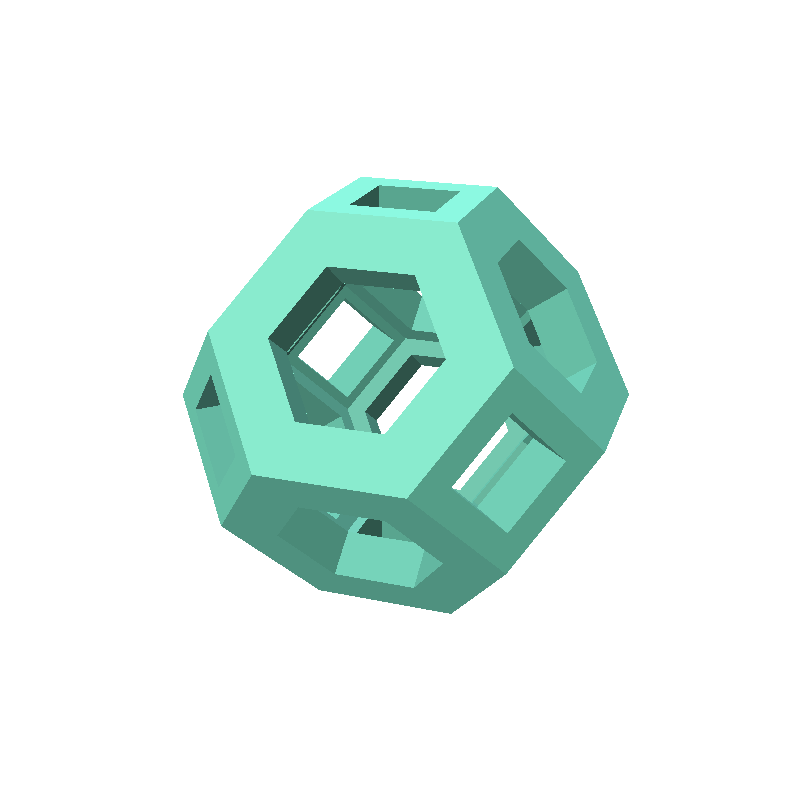}
\caption{Starting geometry. $s_{\mathrm{i}}=0.4\;,\;s_{\mathrm{m}}=0.46\;,\;s_{\mathrm{e}}=0.55\;,\;s_{\mathrm{f}}=0.8$.}
\end{figure}
\begin{figure}[H]
\centering
\includegraphics[scale=0.12]{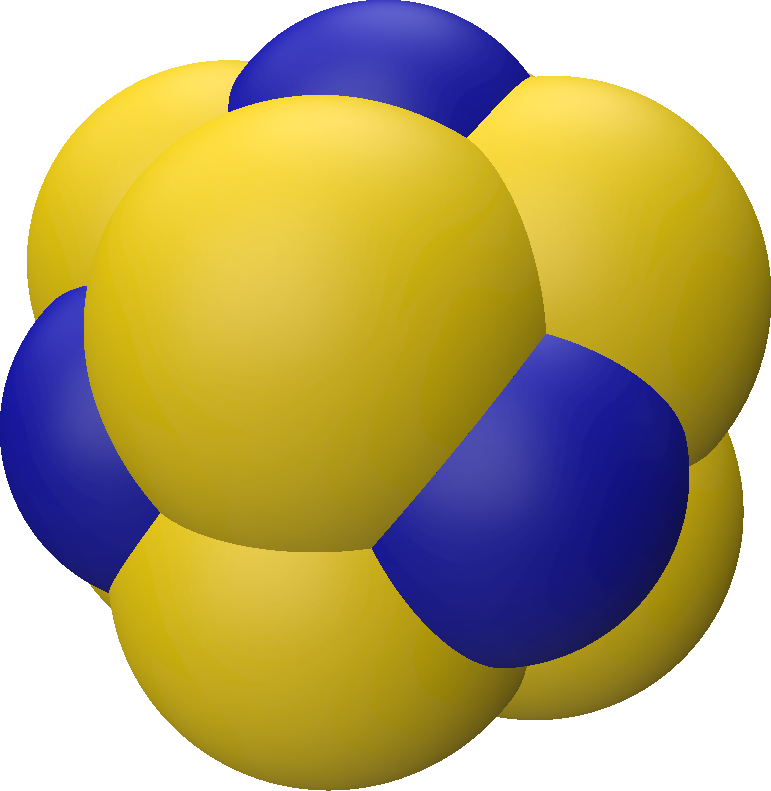}\includegraphics[scale=0.12]{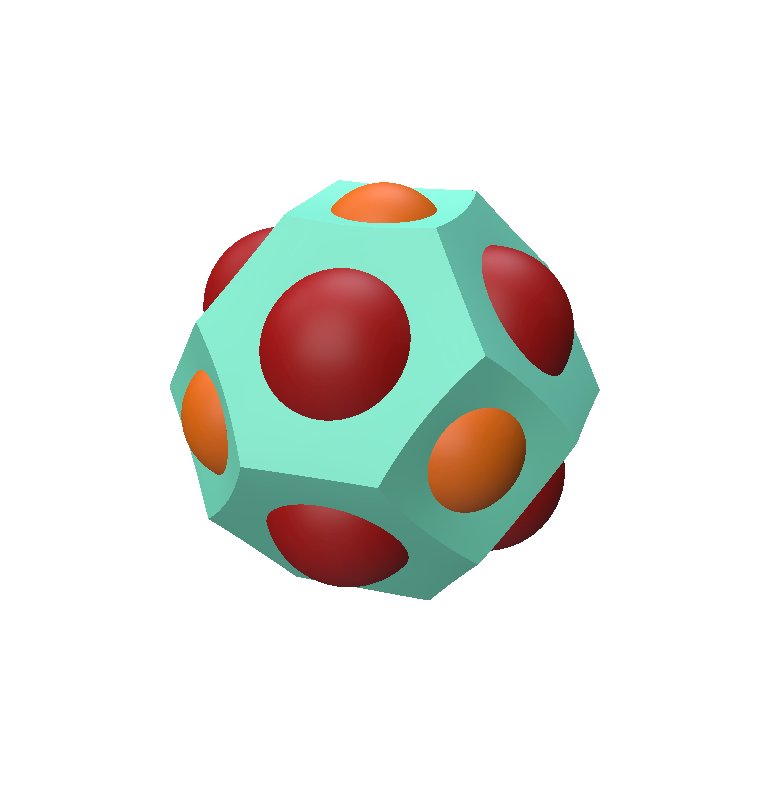}\hspace{-1cm}\includegraphics[scale=0.12]{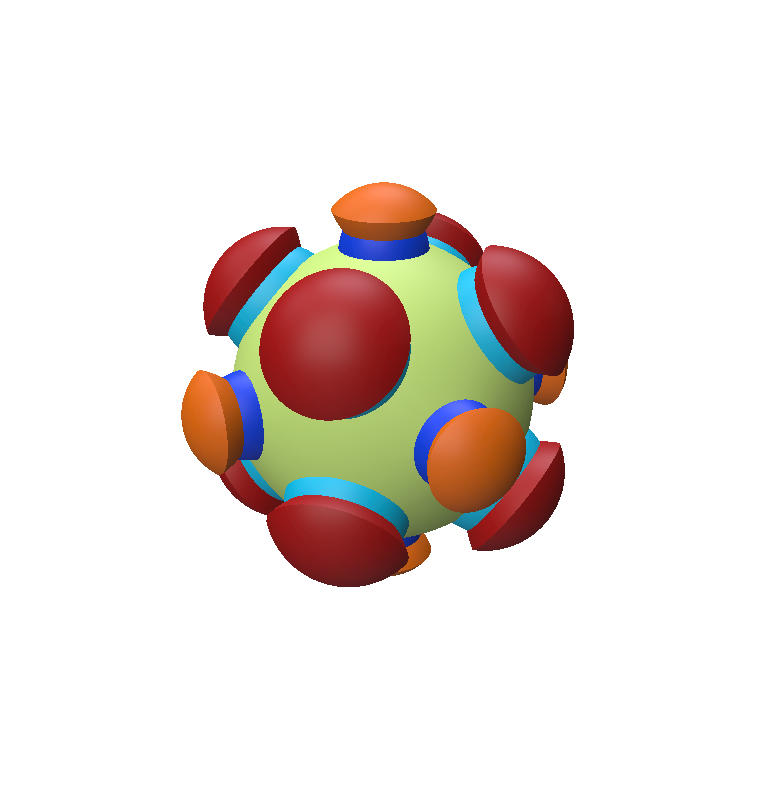}\hspace{-1cm}\includegraphics[scale=0.12]{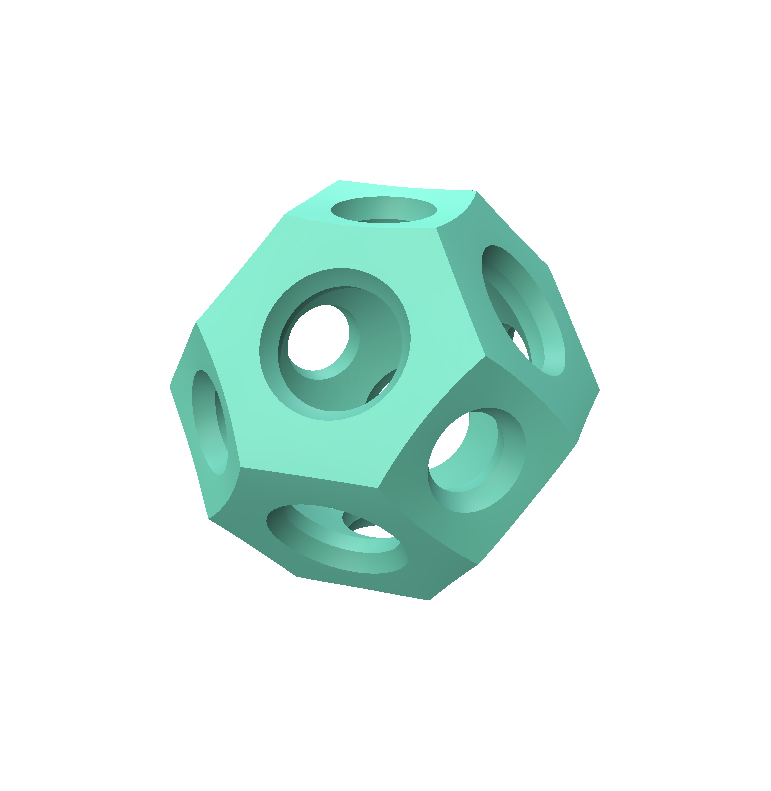}
\caption{Stable cluster. Total area: $54.83$.}
\end{figure}
\begin{table}[H]
\centering
\begin{tabular}{|c|c|c|}
\hline
Discretization & Lowest eigenvalue & Highest eigenvalue\\
\hline
$3\times10^{-2}$ & $1.7\times10^{-5}$ & $14$\\
\hline
$10^{-2}$ & $6.5\times10^{-7}$ & $17$\\
\hline
$8\times10^{-3}$ & $3.0\times10^{-7}$ & $17$\\
\hline
\end{tabular}
\caption{Hessian matrix eigenvalues for the different discretizations.}
\end{table}
\subsection{Great rhombicuboctahedron\\(truncated cuboctahedron)}
The great rhombicuboctahedron has $12$ square faces, $8$ hexagonal faces and $6$ octagonal faces, $72$ edges and $48$ vertices. The cluster has $80$ bubbles and the multiple torus bubble has genus $25$. Vertices coordinates of the great rhombicuboctahedron of unit edge length can be chosen as the permutations of
\begin{equation}
\frac{1}{2}\begin{pmatrix}\pm1\\\pm(1+\sqrt{2})\\\pm(1+2\sqrt{2})\end{pmatrix}.
\end{equation}
\begin{figure}[H]
\centering
\includegraphics[scale=0.1]{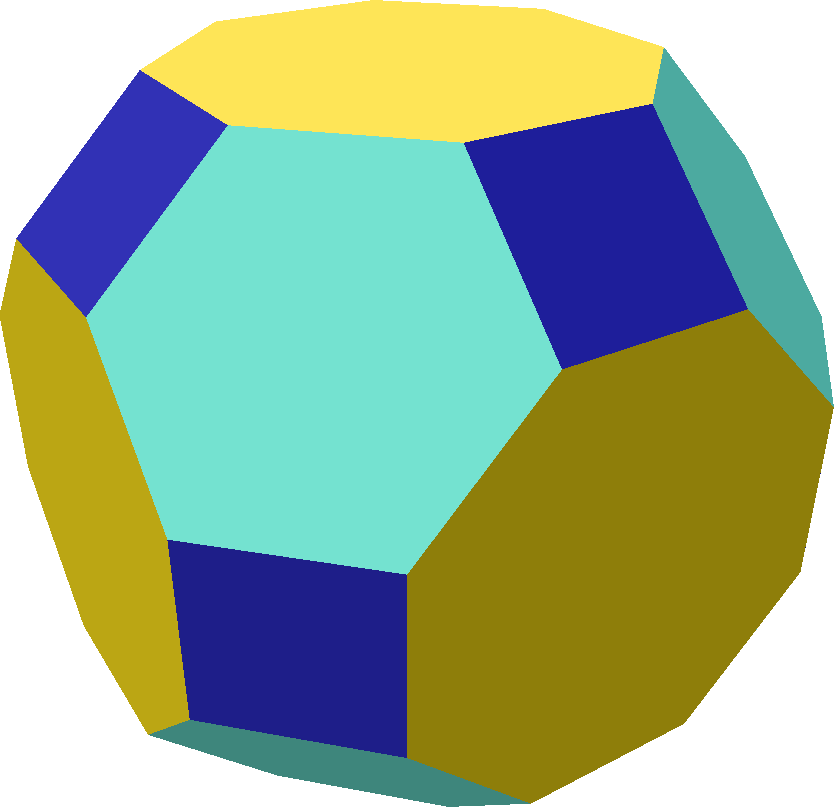}\includegraphics[scale=0.1]{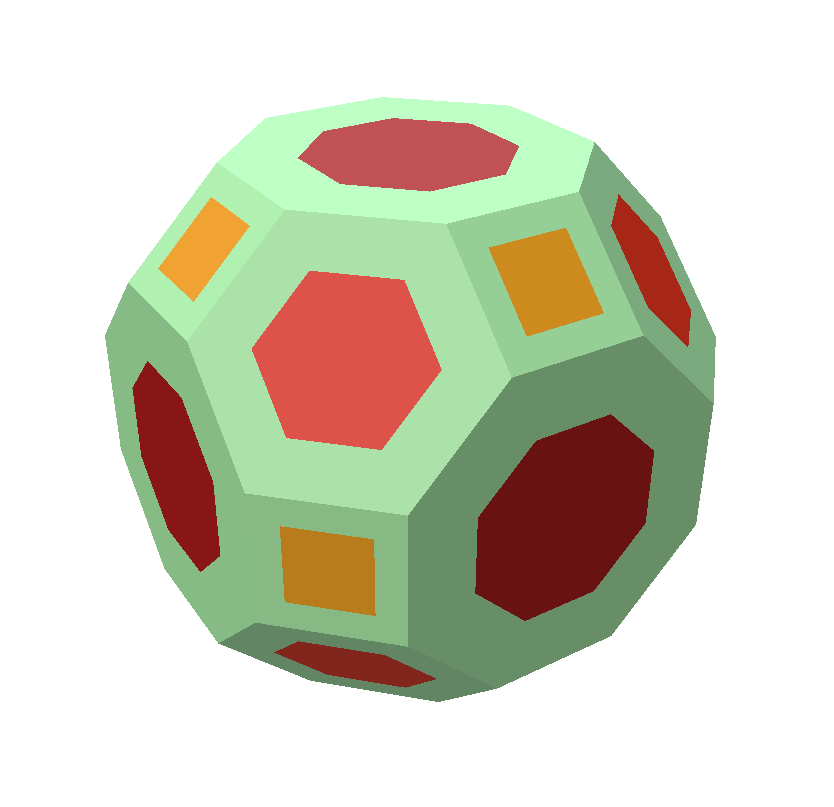}\hspace{-0.6cm}\includegraphics[scale=0.11]{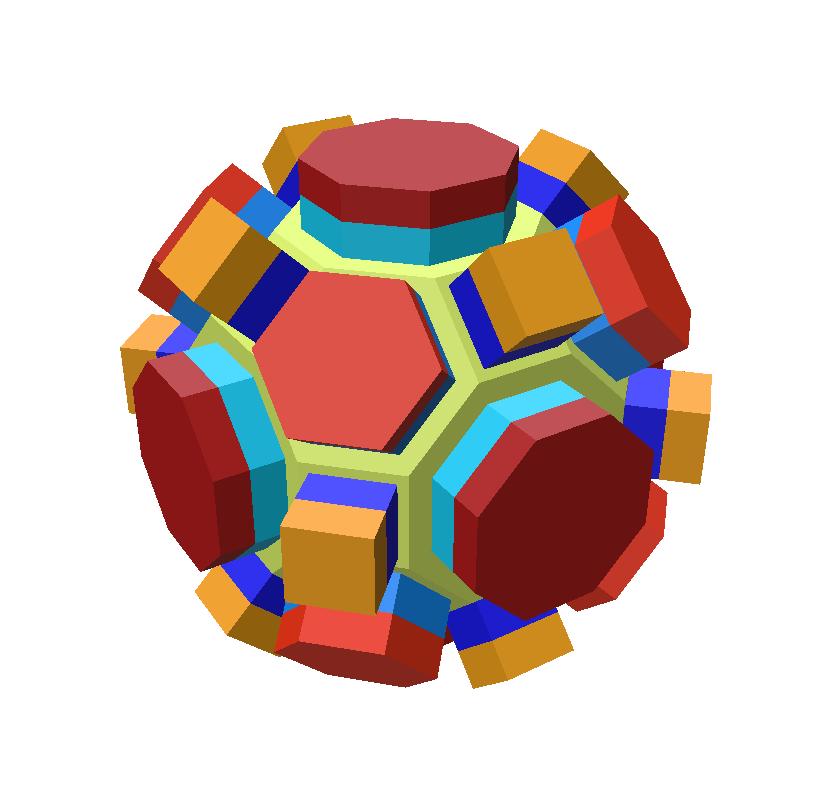}\hspace{-0.6cm}\includegraphics[scale=0.1]{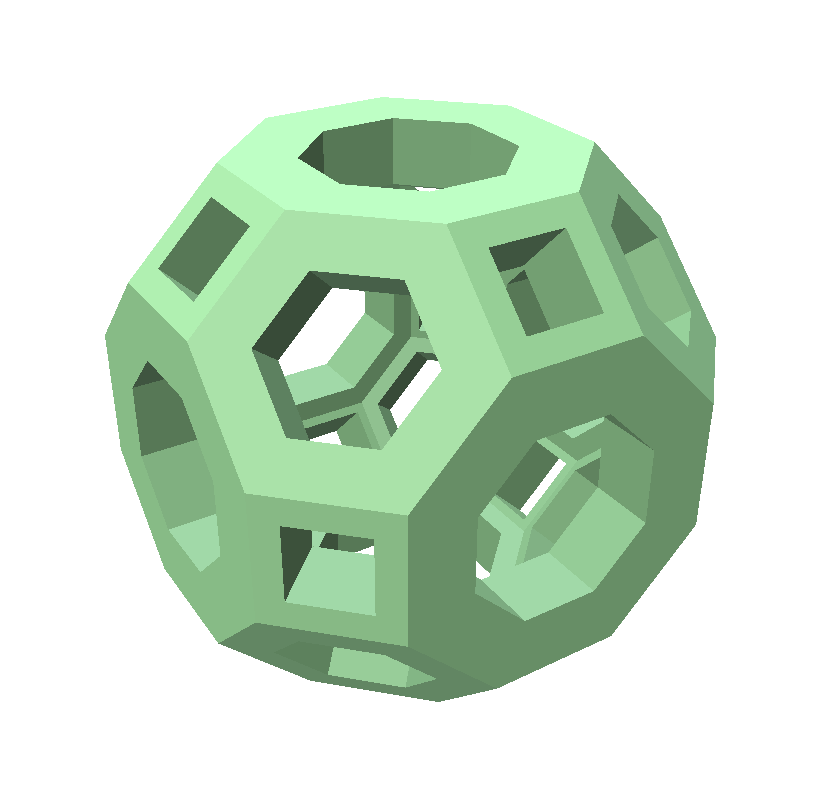}
\caption{Starting geometry. $s_{\mathrm{i}}=0.55\;,\;s_{\mathrm{m}}=0.65\;,\;s_{\mathrm{e}}=0.75\;,\;s_{\mathrm{f}}=0.8$.}
\end{figure}
\begin{figure}[H]
\centering
\includegraphics[scale=0.1]{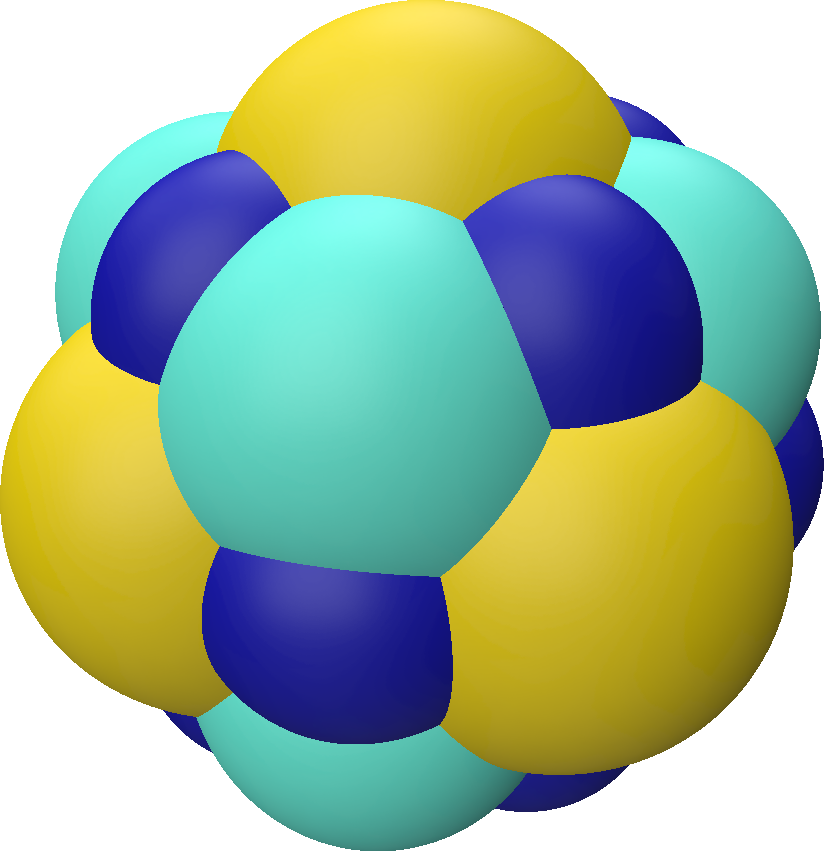}\includegraphics[scale=0.1]{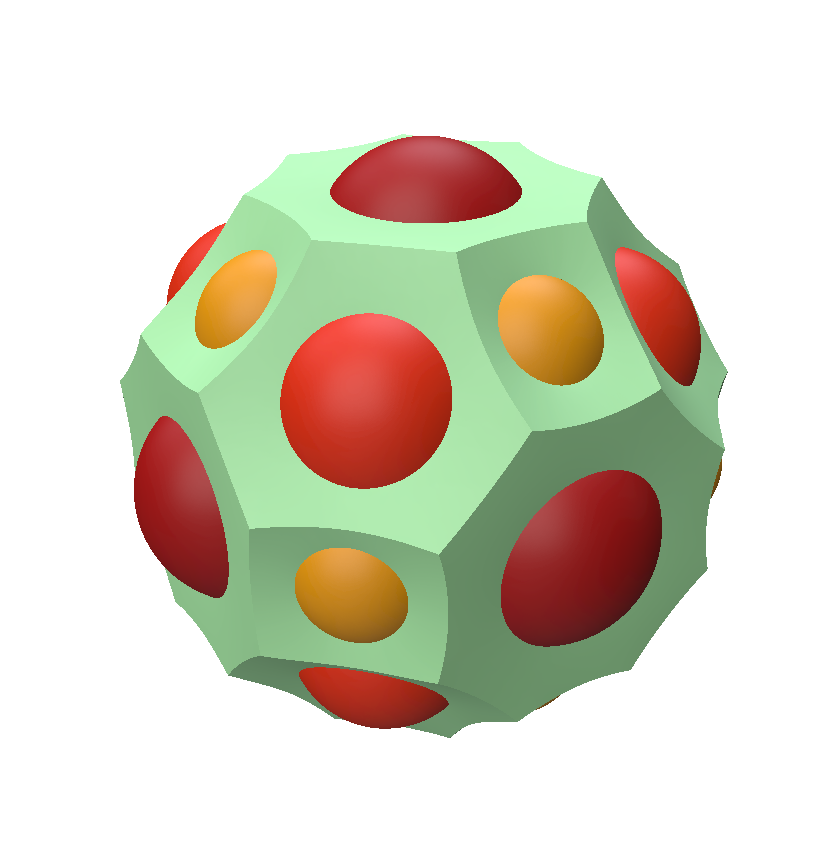}\hspace{-0.6cm}\includegraphics[scale=0.11]{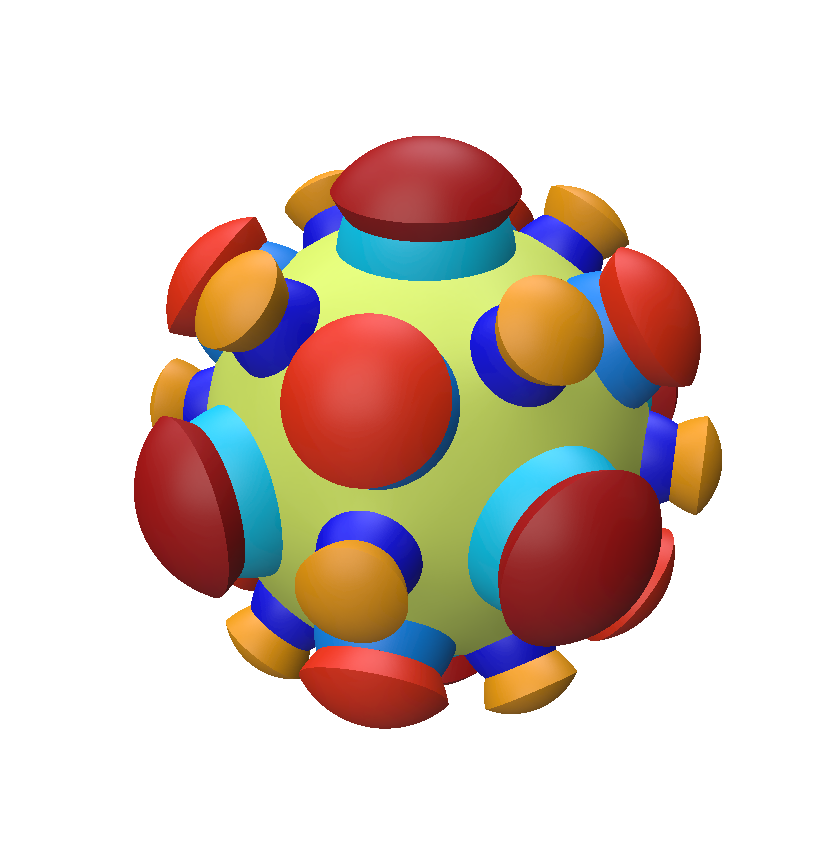}\hspace{-0.6cm}\includegraphics[scale=0.1]{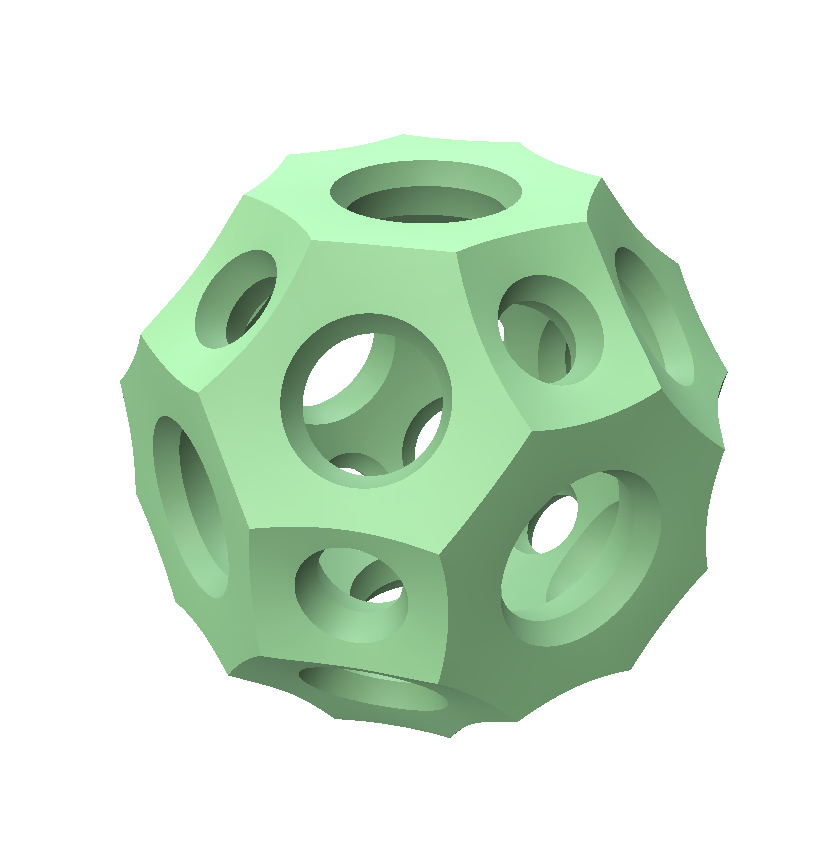}
\caption{Stable cluster. Total area: $165.40$.}
\end{figure}
\begin{table}[H]
\centering
\begin{tabular}{|c|c|c|}
\hline
Discretization & Lowest eigenvalue & Highest eigenvalue\\
\hline
$5\times10^{-2}$ & $5.6\times10^{-5}$ & $11$\\
\hline
$2\times10^{-2}$ & $3.1\times10^{-6}$ & $12$\\
\hline
$10^{-2}$ & $3.8\times10^{-7}$ & $11$\\
\hline
\end{tabular}
\caption{Hessian matrix eigenvalues for the different discretizations.}
\end{table}
\subsection{Truncated dodecahedron}
The truncated dodecahedron has $20$ triangular faces and $12$ decagonal faces, $90$ edges and $60$ vertices. The cluster has $98$ bubbles and the multiple torus bubble has genus $31$. Vertices coordinates of the truncated dodecahedron of unit edge length can be chosen as the cyclic permutations of
\begin{equation}
\frac{1}{2}\begin{pmatrix}0\\\pm1\\\pm(3\varphi+1)\end{pmatrix},
\frac{1}{2}\begin{pmatrix}\pm1\\\pm(\varphi+1)\\\pm2(\varphi+1)\end{pmatrix},
\frac{1}{2}\begin{pmatrix}\pm(\varphi+1)\\\pm2\varphi\\\pm(2\varphi+1)\end{pmatrix}\quad,\quad\varphi=\frac{1+\sqrt{5}}{2}.
\end{equation}
\begin{figure}[H]
\centering
\includegraphics[scale=0.12]{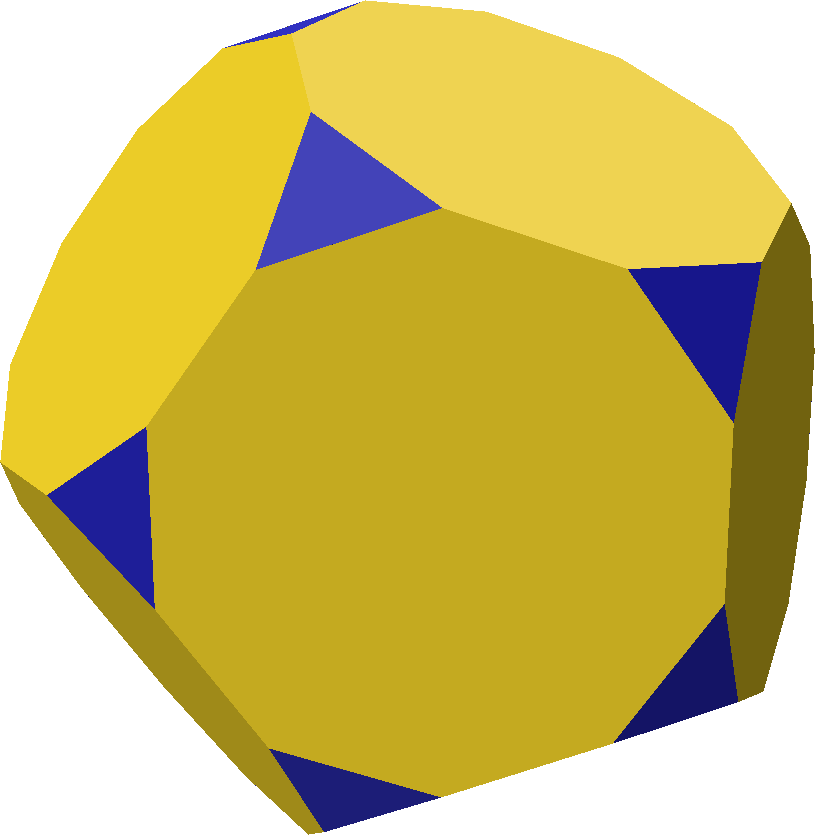}\hspace{-0.2cm}\includegraphics[scale=0.12]{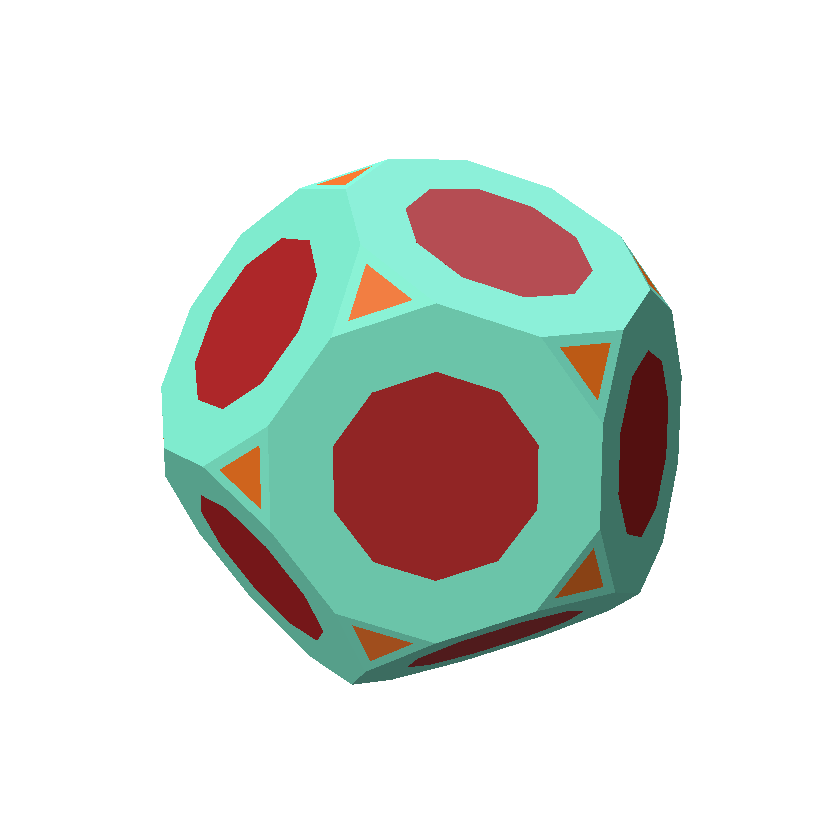}\hspace{-1cm}\includegraphics[scale=0.12]{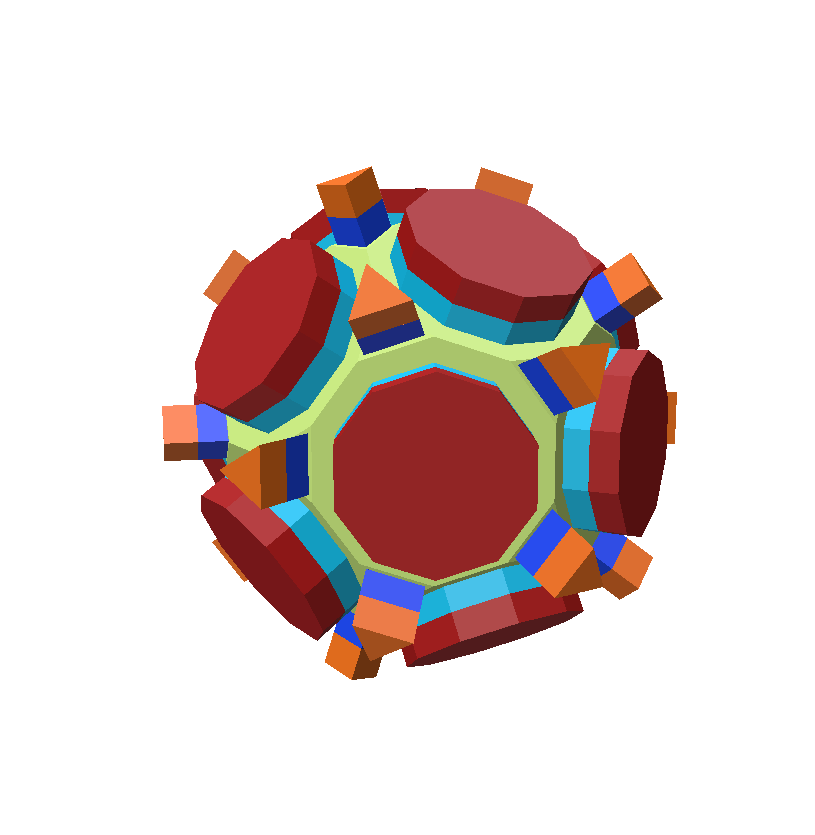}\hspace{-1cm}\includegraphics[scale=0.12]{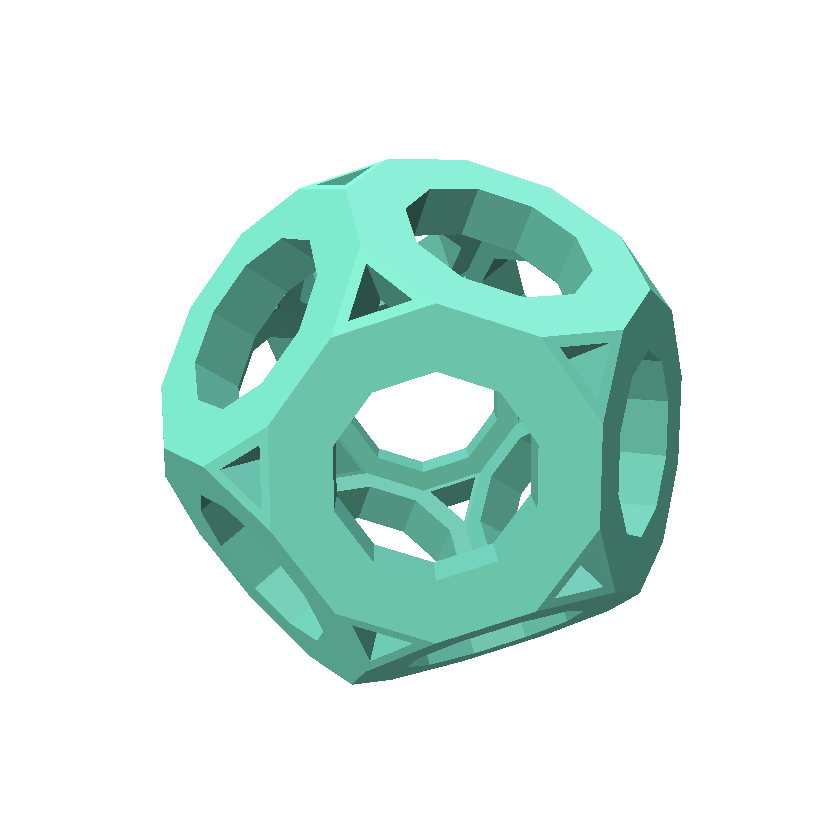}
\caption{Starting geometry. $s_{\mathrm{i}}=0.5\;,\;s_{\mathrm{m}}=0.57\;,\;s_{\mathrm{e}}=0.65\;,\;s_{\mathrm{f}}=0.8$.}
\end{figure}
\begin{figure}[H]
\centering
\includegraphics[scale=0.12]{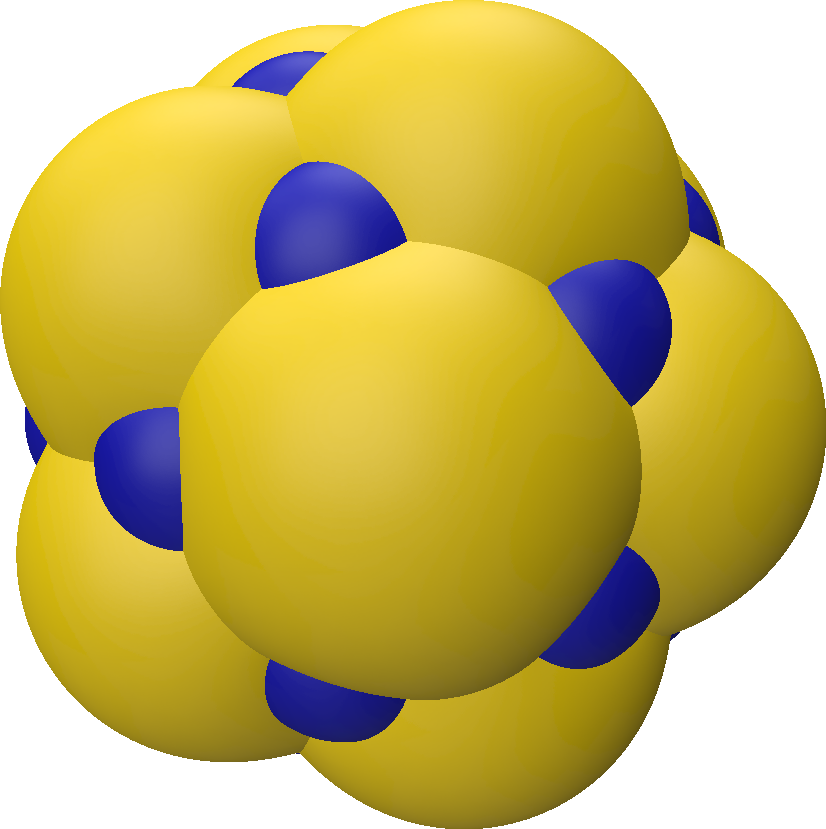}\includegraphics[scale=0.12]{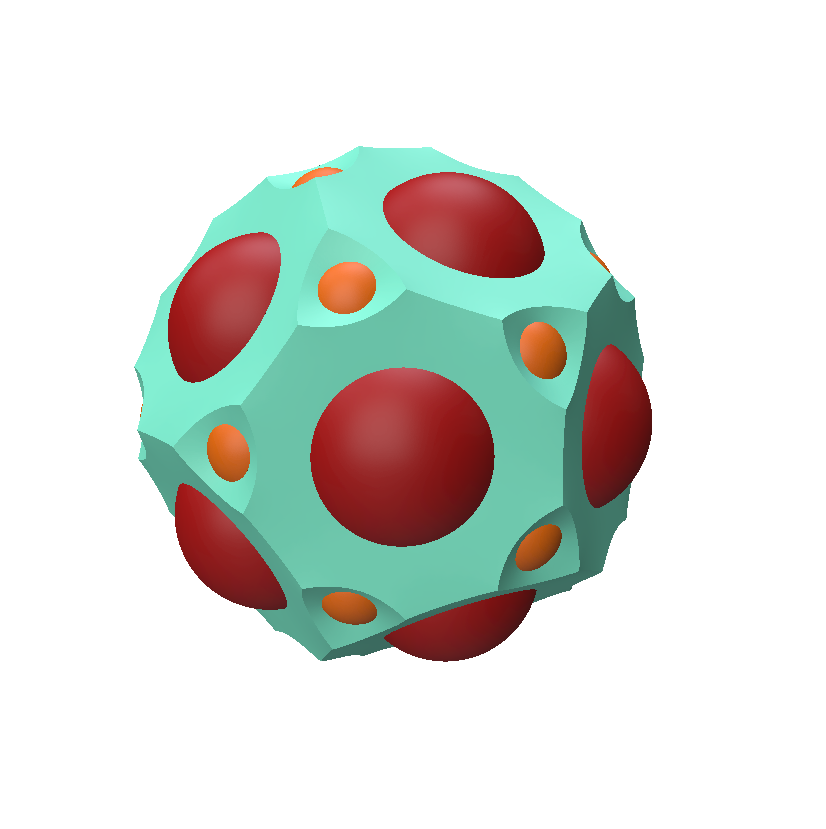}\hspace{-1cm}\includegraphics[scale=0.12]{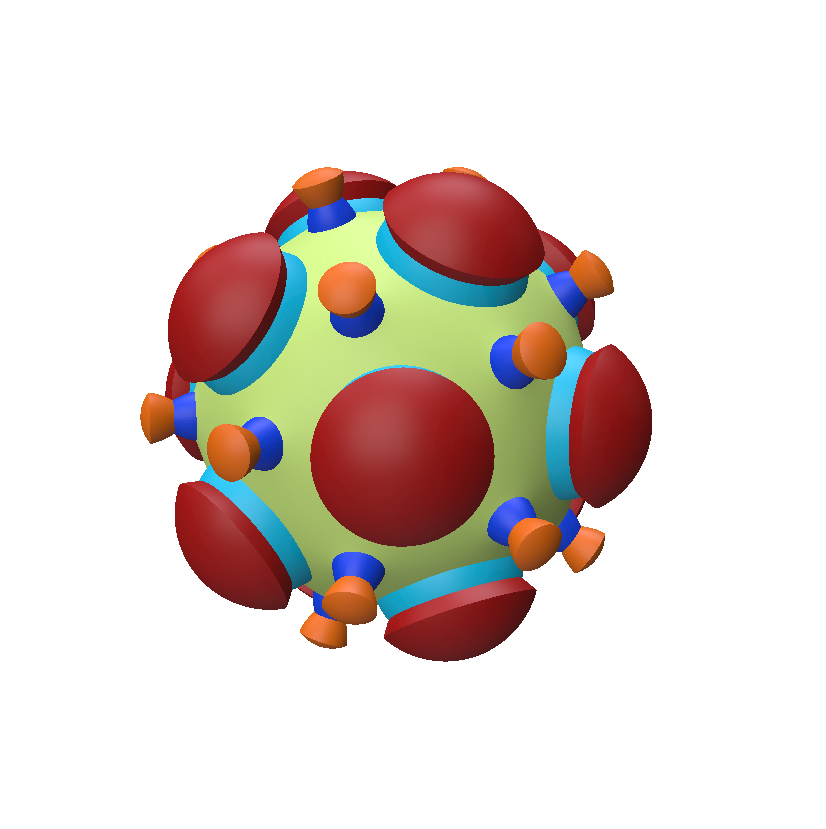}\hspace{-1cm}\includegraphics[scale=0.12]{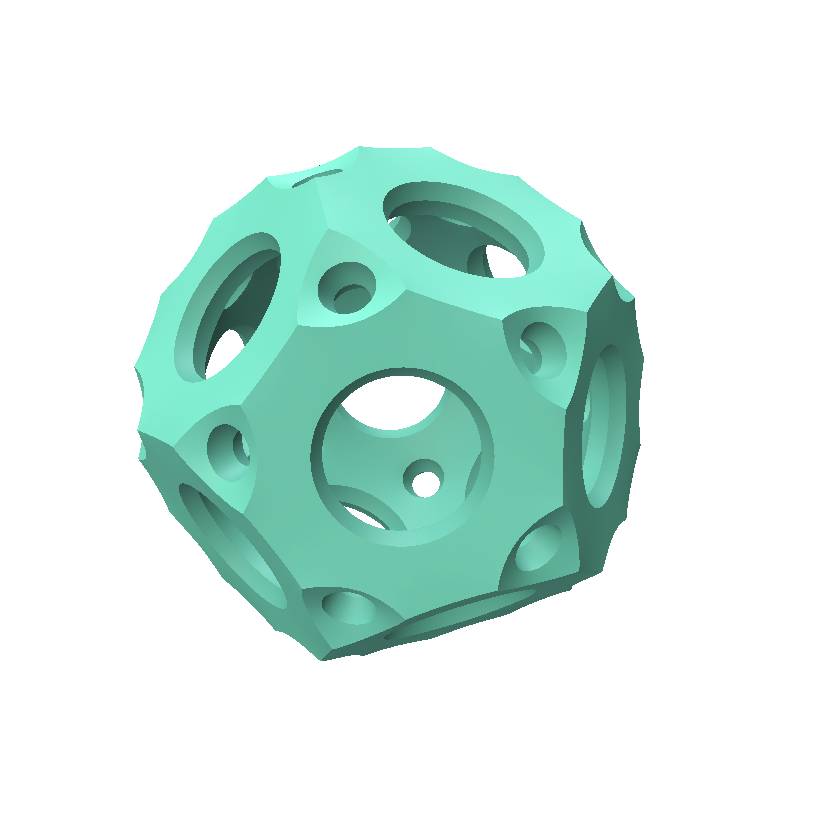}
\caption{Stable cluster. Total area: $245.43$.}
\end{figure}
\begin{table}[H]
\centering
\begin{tabular}{|c|c|c|}
\hline
Discretization & Lowest eigenvalue & Highest eigenvalue\\
\hline
$5\times10^{-2}$ & $5.3\times10^{-5}$ & $14$\\
\hline
$3\times10^{-2}$ & $1.0\times10^{-5}$ & $14$\\
\hline
$1.5\times10^{-2}$ & $1.4\times10^{-6}$ & $14$\\
\hline
\end{tabular}
\caption{Hessian matrix eigenvalues for the different discretizations.}
\end{table}
\subsection{Truncated icosahedron}
The truncated icosahedron has $12$ pentagonal faces and $20$ hexagonal faces, $90$ edges and $60$ vertices. The cluster has $98$ bubbles and the multiple torus bubble has genus $31$. Vertices coordinates of the truncated icosahedron of unit edge length can be chosen as the cyclic permutations of
\begin{equation}
\frac{1}{2}\begin{pmatrix}0\\\pm1\\\pm3\varphi\end{pmatrix},
\frac{1}{2}\begin{pmatrix}\pm2\\\pm(2\varphi+1)\\\pm\varphi\end{pmatrix},
\frac{1}{2}\begin{pmatrix}\pm1\\\pm(\varphi+2)\\\pm2\varphi\end{pmatrix}\quad,\quad\varphi=\frac{1+\sqrt{5}}{2}.
\end{equation}
\begin{figure}[H]
\centering
\includegraphics[scale=0.1]{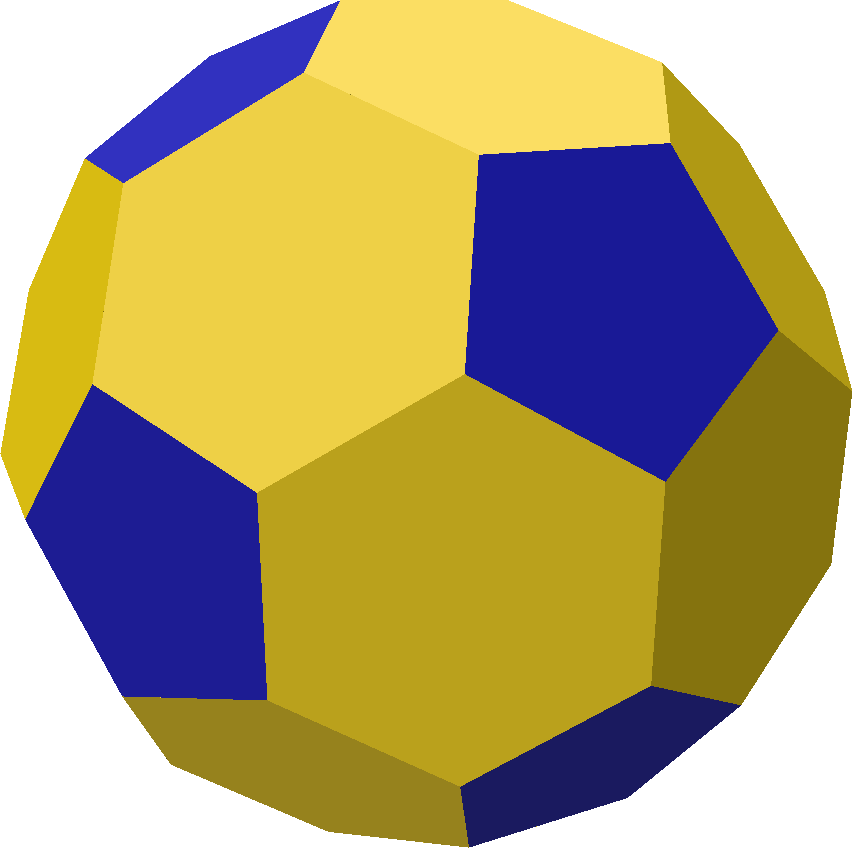}\includegraphics[scale=0.1]{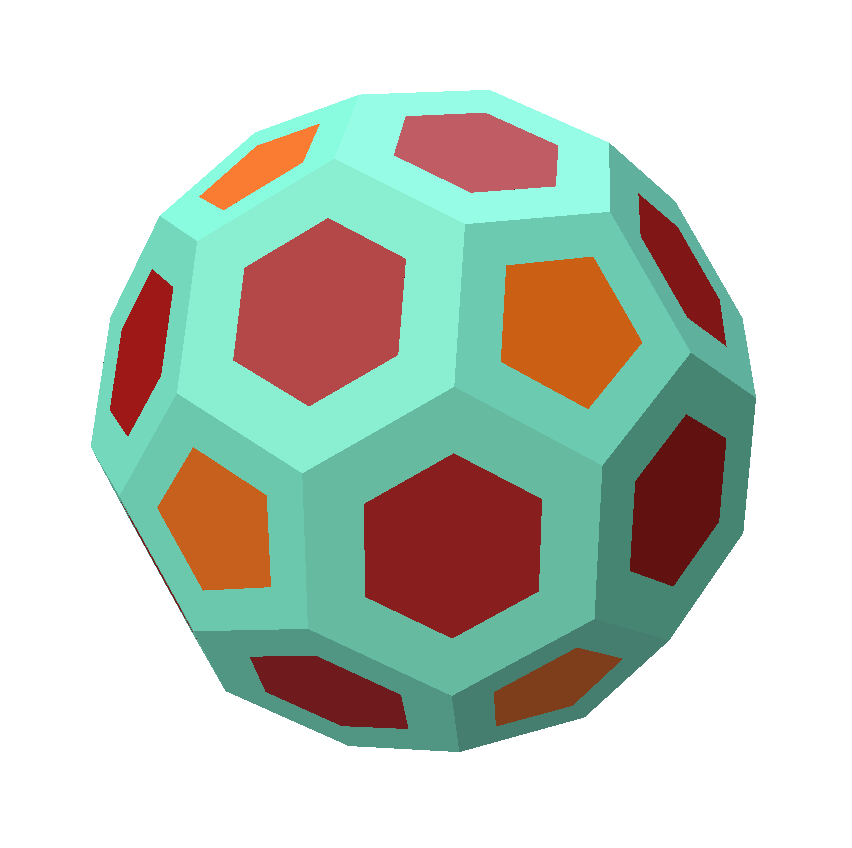}\hspace{-0.5cm}\includegraphics[scale=0.1]{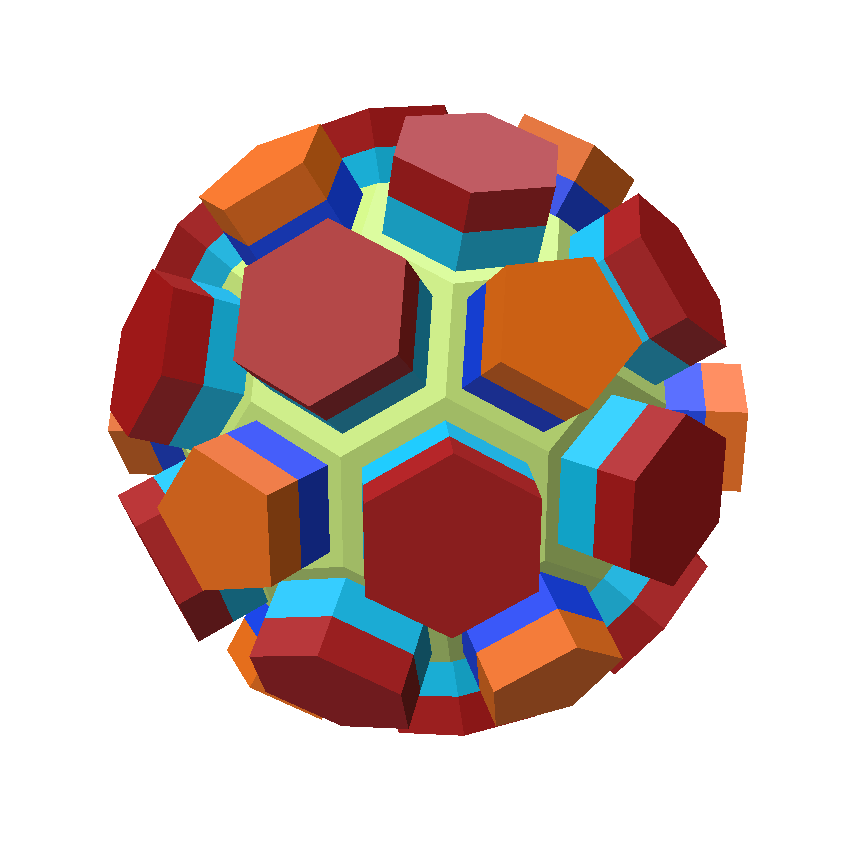}\hspace{-0.5cm}\includegraphics[scale=0.1]{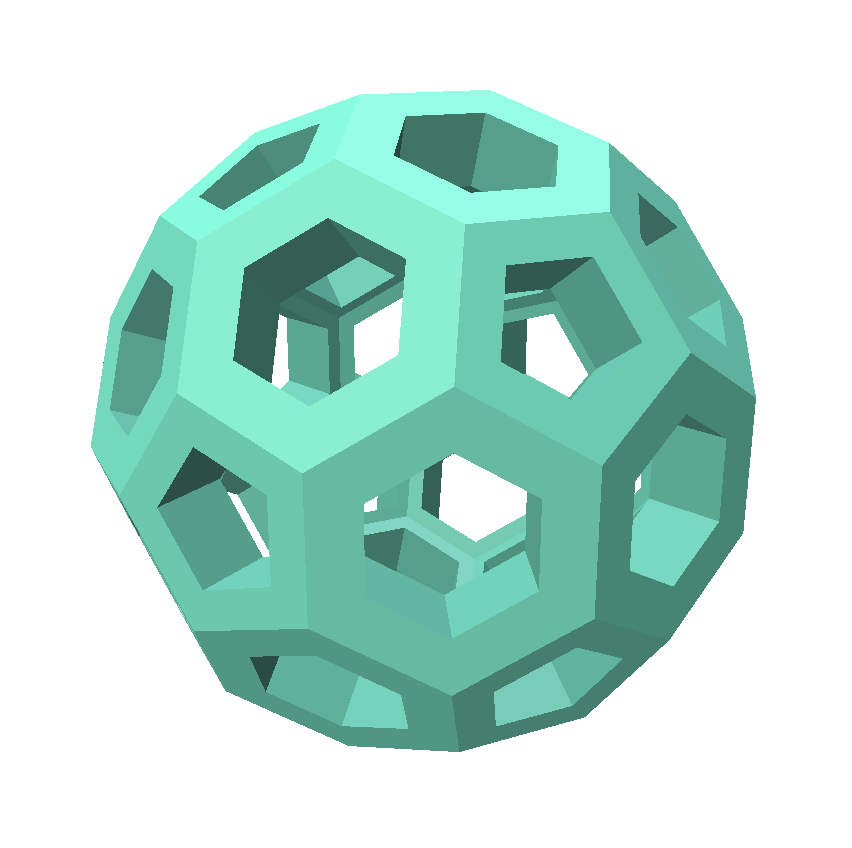}
\caption{Starting geometry. $s_{\mathrm{i}}=0.6\;,\;s_{\mathrm{m}}=0.7\;,\;s_{\mathrm{e}}=0.8\;,\;s_{\mathrm{f}}=0.8$.}
\end{figure}
\begin{figure}[H]
\centering
\includegraphics[scale=0.1]{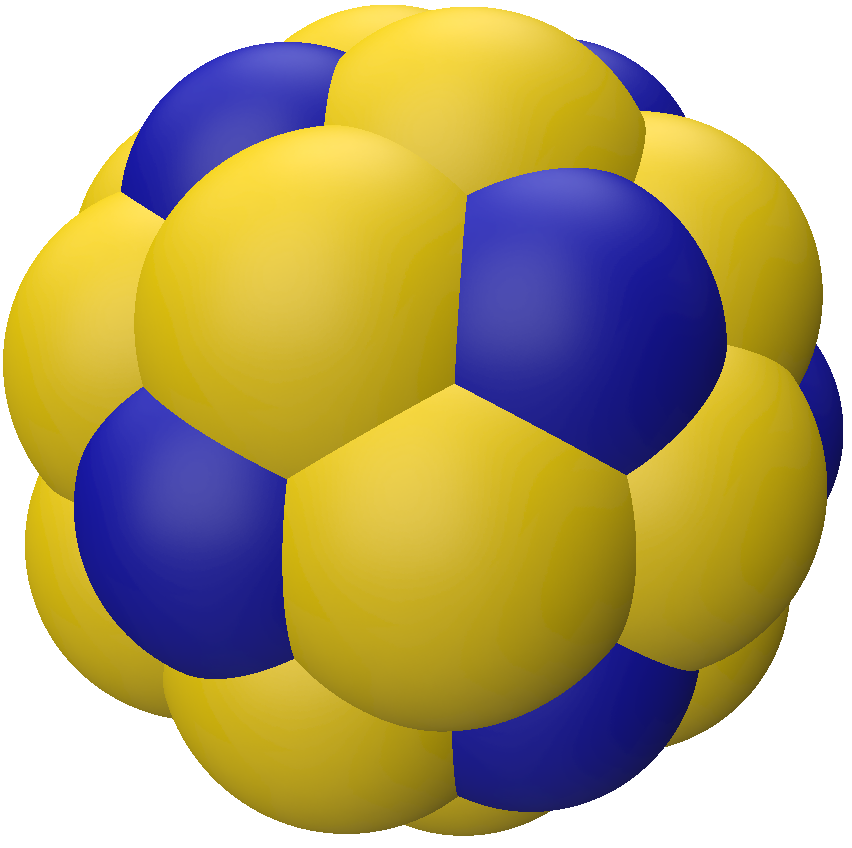}\includegraphics[scale=0.1]{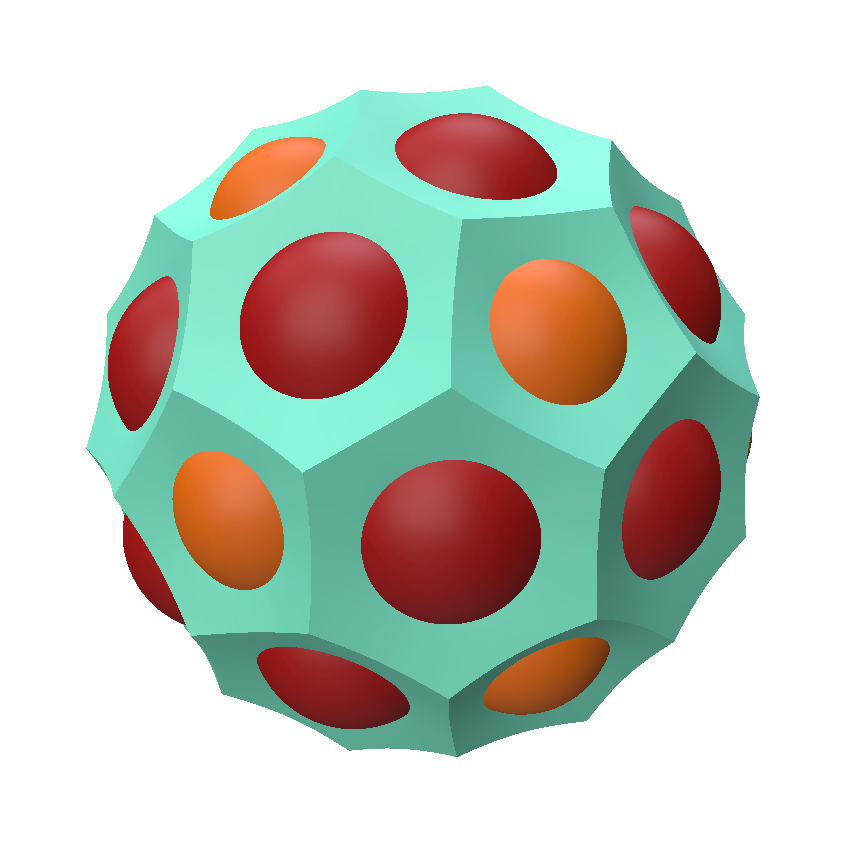}\hspace{-0.4cm}\includegraphics[scale=0.1]{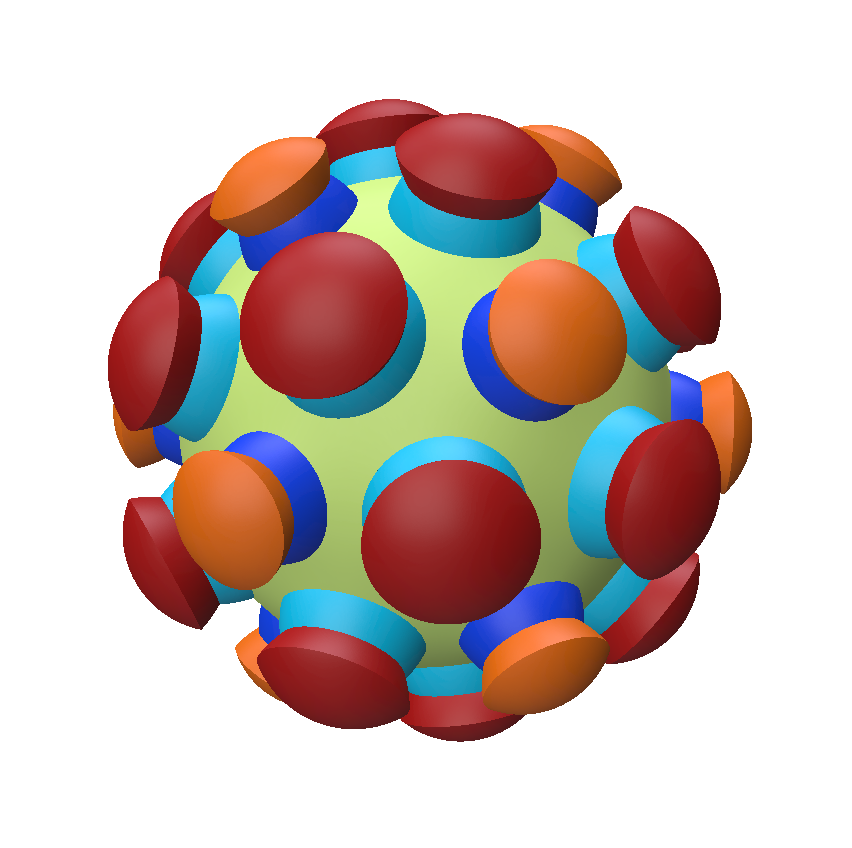}\hspace{-0.4cm}\includegraphics[scale=0.1]{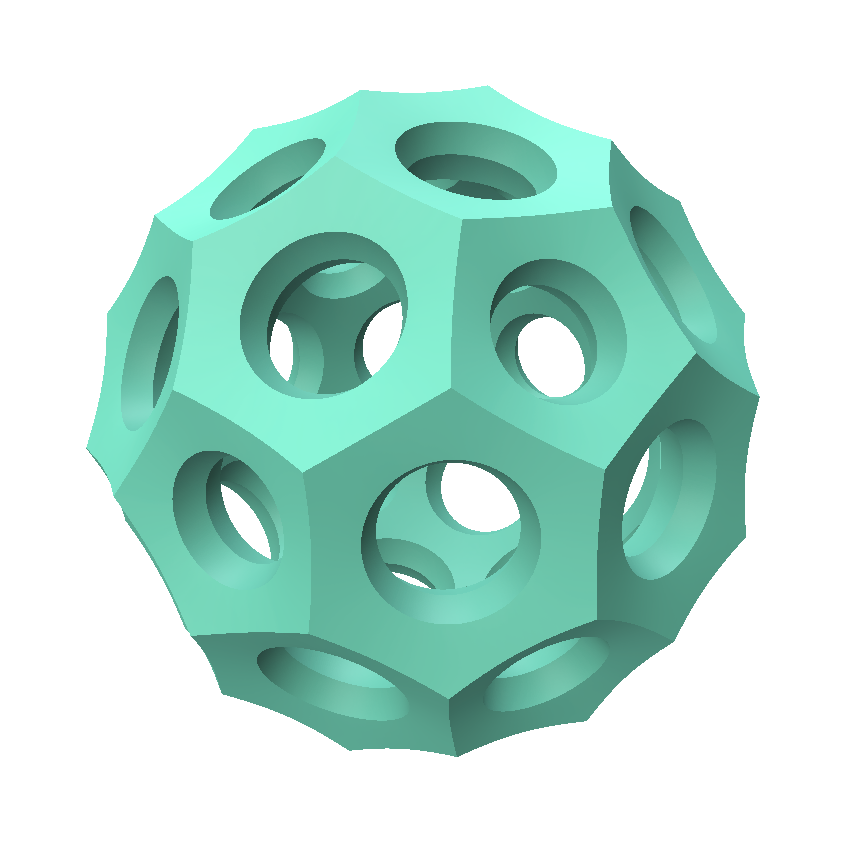}
\caption{Stable cluster. Total area: $216.12$.}
\end{figure}
\begin{table}[H]
\centering
\begin{tabular}{|c|c|c|}
\hline
Discretization & Lowest eigenvalue & Highest eigenvalue\\
\hline
$6\times10^{-2}$ & $8.2\times10^{-5}$ & $16$\\
\hline
$3\times10^{-2}$ & $1.1\times10^{-5}$ & $16$\\
\hline
$1.5\times10^{-2}$ & $1.2\times10^{-6}$ & $17$\\
\hline
\end{tabular}
\caption{Hessian matrix eigenvalues for the different discretizations.}
\end{table}
\subsection{Great rhombicosidodecahedron\\(truncated icosidodecahedron)}
The great rhombicosidodecahedron has $30$ square faces, $20$ hexagonal faces and $12$ decagonal faces, $180$ edges and $120$ vertices. The cluster has $188$ bubbles and the multiple torus bubble has genus $61$. Vertices coordinates of the great rhombicosidodecahedron of unit edge length can be chosen as the cyclic permutations of
\begin{equation}
\begin{split}
\frac{1}{2}\begin{pmatrix}\pm1\\\pm1\\\pm(4\varphi+1)\end{pmatrix},
\frac{1}{2}\begin{pmatrix}\pm2\\\pm(\varphi+1)\\\pm(3\varphi+2)\end{pmatrix},
\frac{1}{2}\begin{pmatrix}\pm1\\\pm(2\varphi+1)\\\pm(2\varphi+3)\end{pmatrix},&\\
\frac{1}{2}\begin{pmatrix}\pm(\varphi+2)\\\pm2\varphi\\\pm(3\varphi+1)\end{pmatrix},
\frac{1}{2}\begin{pmatrix}\pm(\varphi+1)\\\pm3\varphi\\\pm2(\varphi+1)\end{pmatrix}\quad,\quad\varphi=\frac{1+\sqrt{5}}{2}.&
\end{split}
\end{equation}
\begin{figure}[H]
\centering
\includegraphics[scale=0.1]{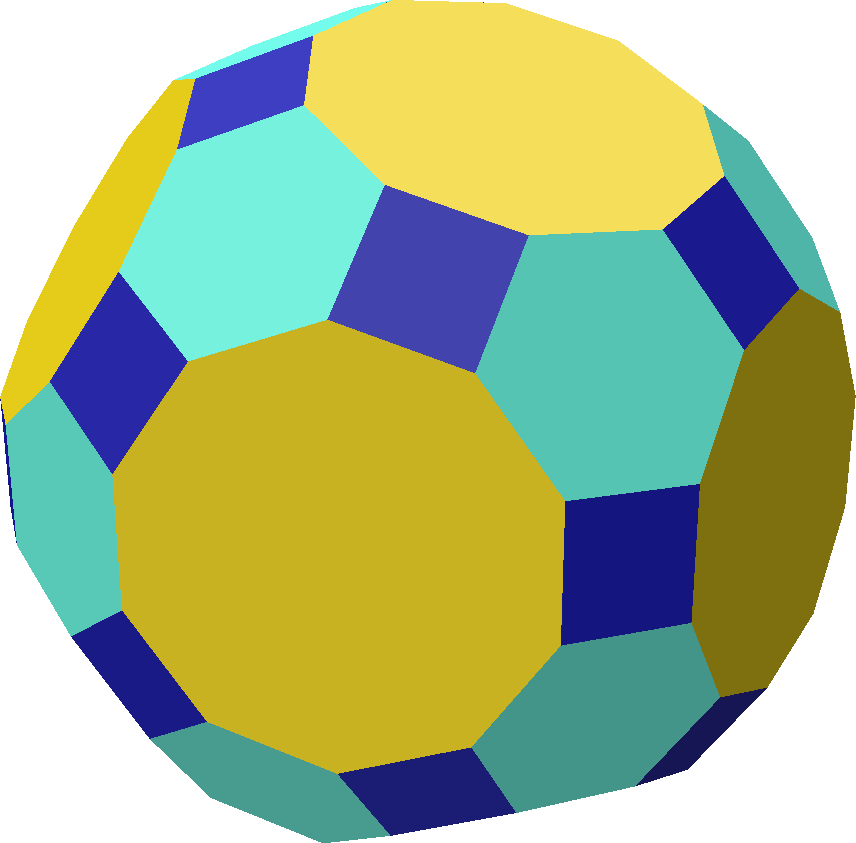}\hspace{-0.2cm}\includegraphics[scale=0.1]{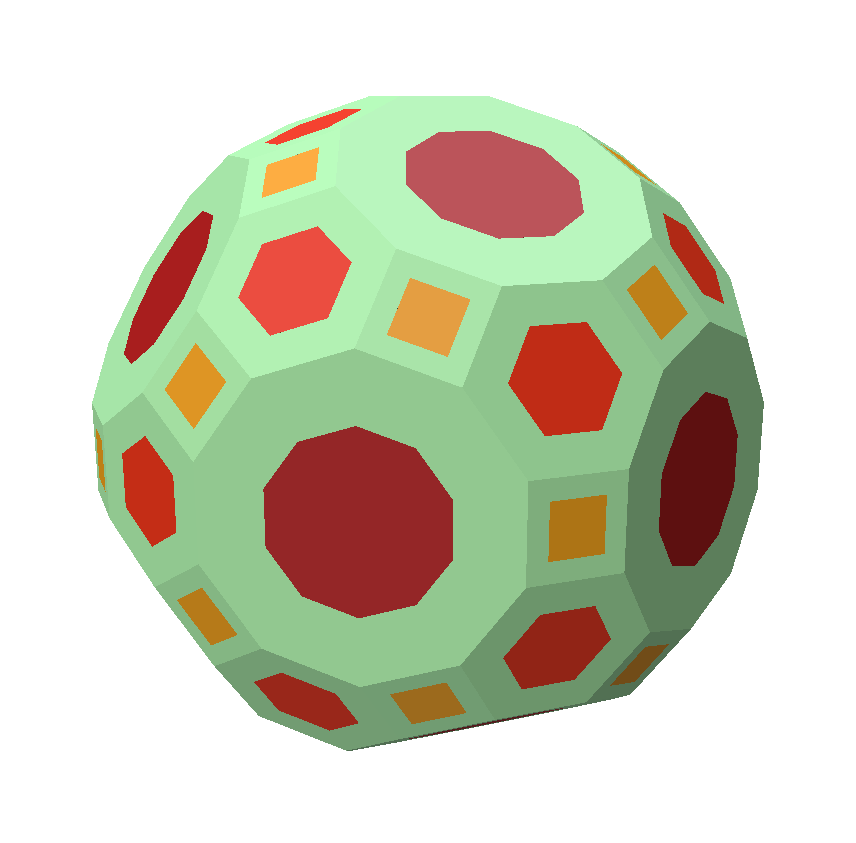}\hspace{-0.5cm}\includegraphics[scale=0.1]{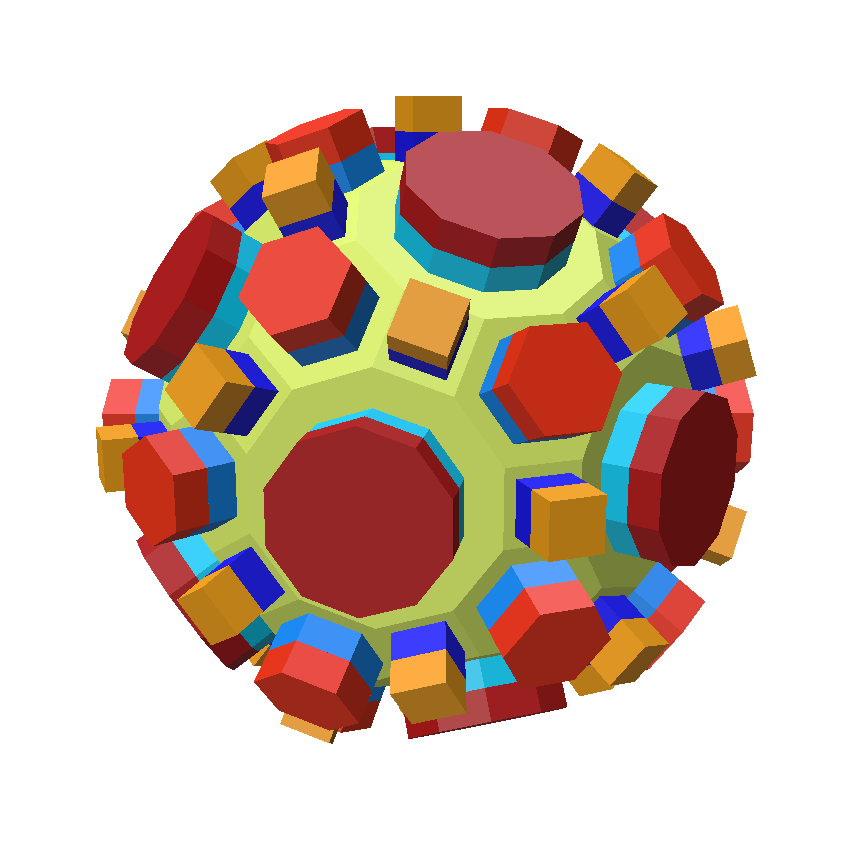}\hspace{-0.5cm}\includegraphics[scale=0.1]{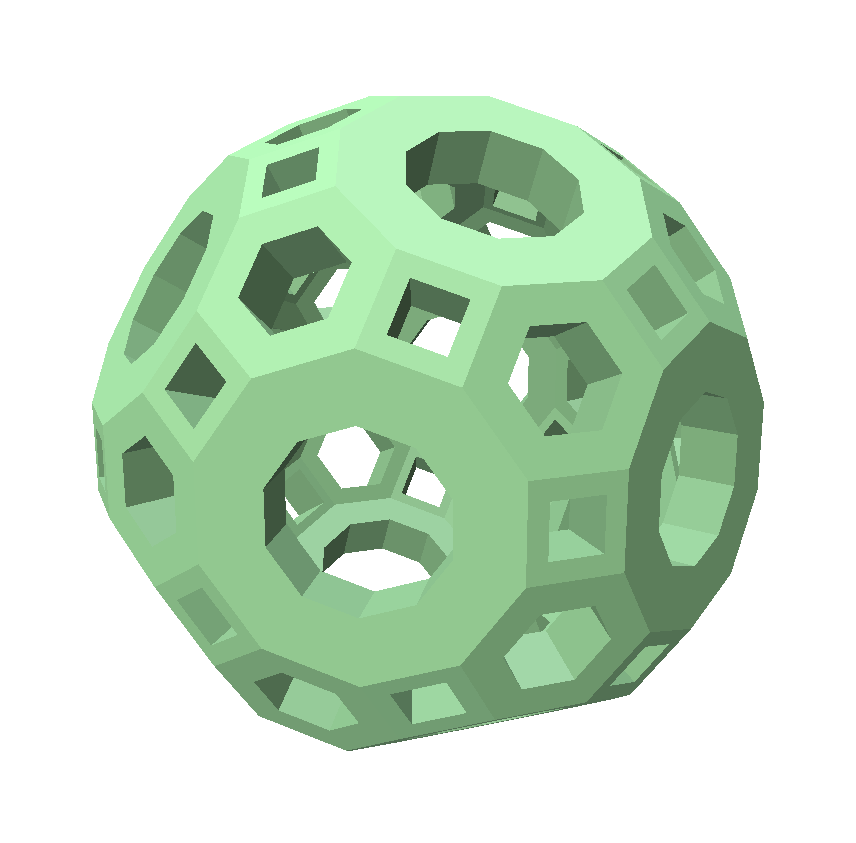}
\caption{Starting geometry. $s_{\mathrm{i}}=0.65\;,\;s_{\mathrm{m}}=0.72\;,\;s_{\mathrm{e}}=0.8\;,\;s_{\mathrm{f}}=0.7$.}
\end{figure}
\begin{figure}[H]
\centering
\includegraphics[scale=0.1]{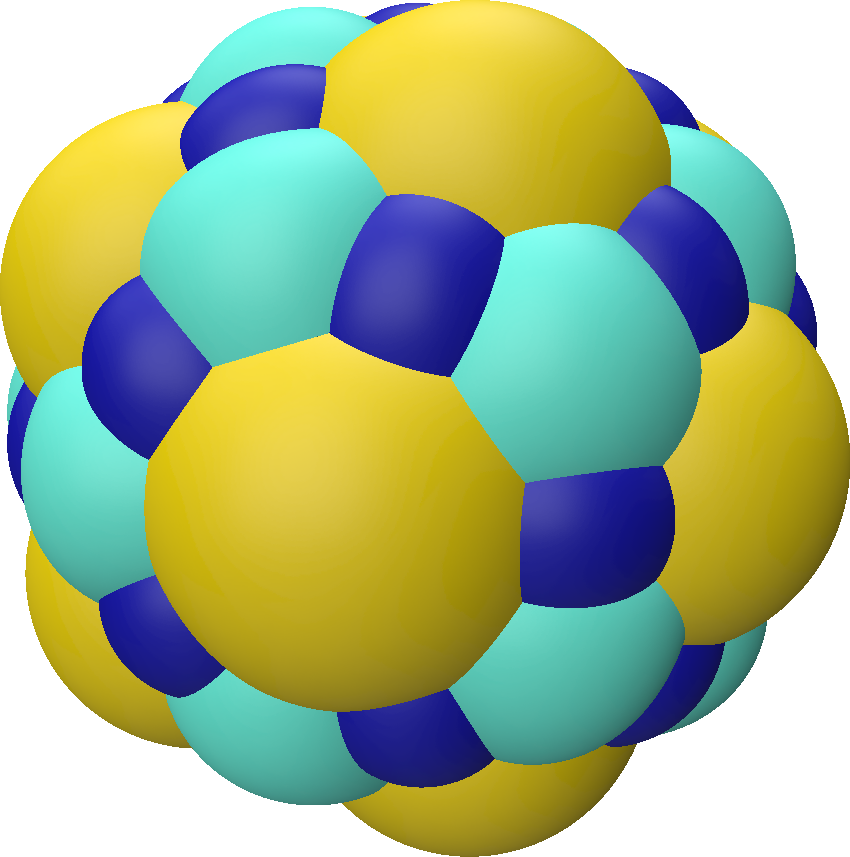}\includegraphics[scale=0.1]{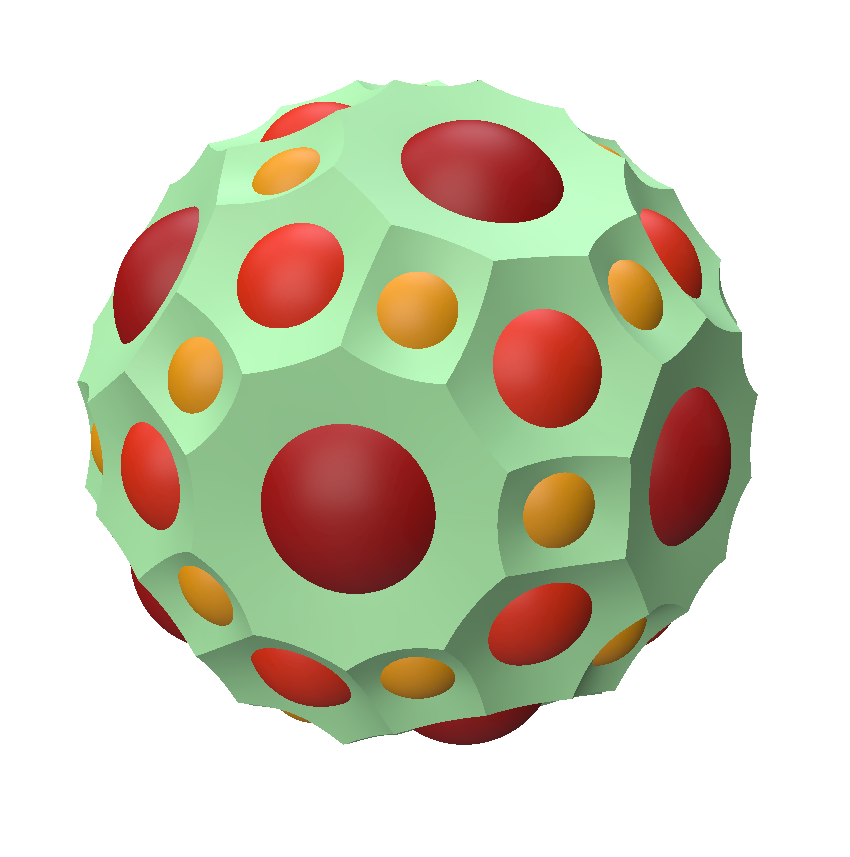}\hspace{-0.2cm}\includegraphics[scale=0.1]{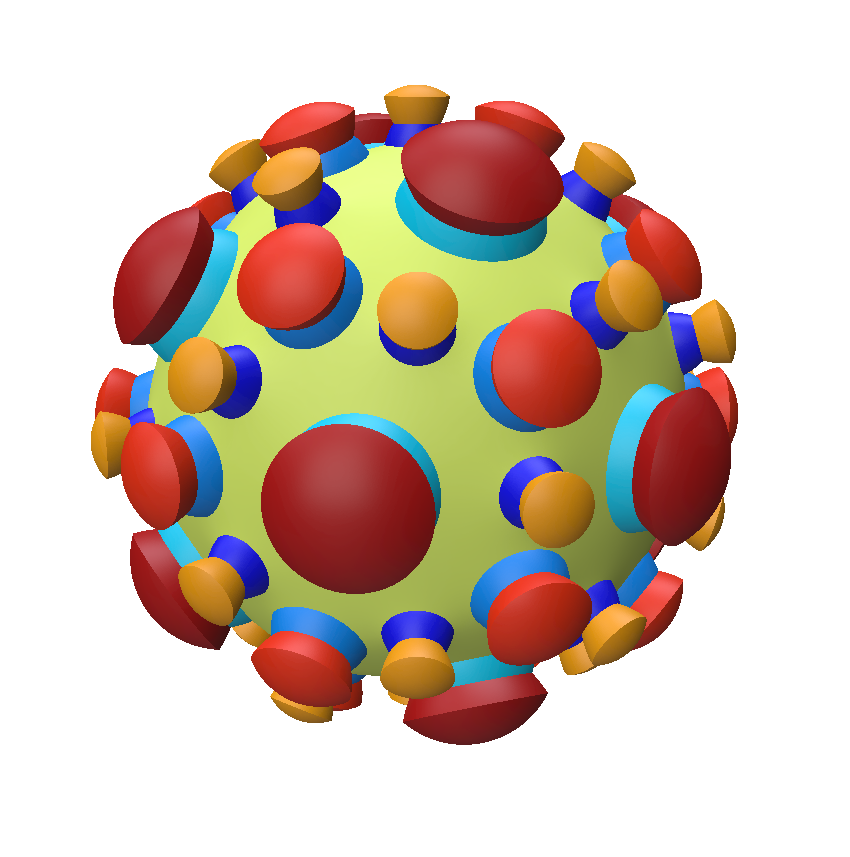}\hspace{-0.2cm}\includegraphics[scale=0.1]{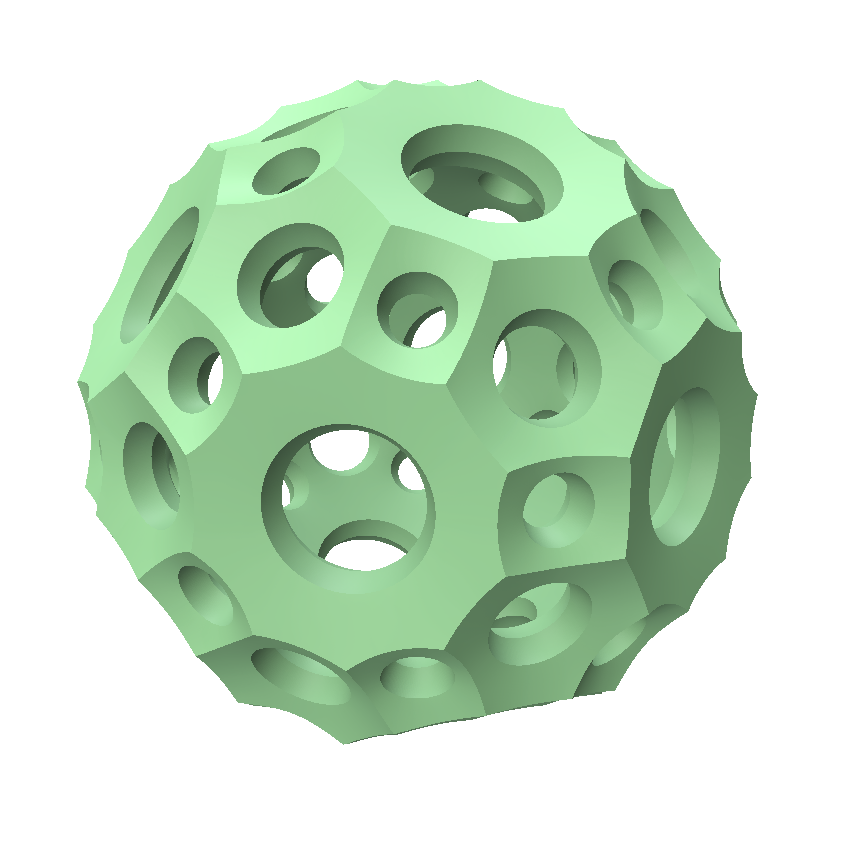}
\caption{Stable cluster. Total area: $545.74$.}
\end{figure}
\begin{table}[H]
\centering
\begin{tabular}{|c|c|c|}
\hline
Discretization & Lowest eigenvalue & Highest eigenvalue\\
\hline
$8\times10^{-2}$ & $8.7\times10^{-5}$ & $11$\\
\hline
$4\times10^{-2}$ & $1.2\times10^{-5}$ & $13$\\
\hline
$2\times10^{-2}$ & $1.4\times10^{-6}$ & $17$\\
\hline
\end{tabular}
\caption{Hessian matrix eigenvalues for the different discretizations.}
\end{table}
\section{Data files}
All simulations results as well as Surface Evolver configuration files and the Python script allowing to create the configurations with some flexibility are available on \href{https://doi.org/10.5281/zenodo.18442419}{Zenodo}. An independent viewer using Polyscope \cite{polyscope} and multimaterial meshes versions (as used by Multitracker \cite{multitracker}) of Surface Evolver dump files are also available. An additional Python script can be used to define more general starting configurations using any convex polyhedron as in the figure \ref{modified_initial_configuration}.
\section{Conclusion}
Using the same construction as in \cite{delbary} for Platonic solids, it has be shown that one can build stable soap bubble clusters with multiple torus bubbles using geometries based on prisms and Archimedean solids as well. One can wonder if the use of inner double bubbles as ``linkers'' could be somehow generalized and used to build more complex configurations.
\bibliography{genus_7_13_25_31_61}
\end{document}